\documentclass[12pt,preprint]{aastex}
\usepackage{natbib}
\slugcomment{for \emph{Astrophysical Journal}, \today}

\shorttitle{Atmospheric Chemistry in Substellar Objects. III.}

\shortauthors{Visscher, Lodders, \& Fegley}

\begin{document}

\title{Atmospheric Chemistry in Giant Planets, Brown Dwarfs, and Low-Mass Dwarf Stars III. Iron, Magnesium, and Silicon}

\author{Channon Visscher$^{1,2}$, Katharina Lodders$^{1}$, and Bruce Fegley, Jr.$^{1}$}

\affil{$^{1}$Planetary Chemistry Laboratory, Department of Earth \& Planetary Sciences, McDonnell Center for the Space Sciences, Washington University, St. Louis, MO 63130-4899}

\affil{$^{2}${Current Address}: Lunar and Planetary Institute, USRA, Houston, TX 77058-1113}

 \email{visscher@lpi.usra.edu, lodders@wustl.edu, bfegley@wustl.edu}

\begin{abstract}
We use thermochemical equilibrium calculations to model iron, magnesium, and silicon chemistry in the atmospheres of giant planets, brown dwarfs, extrasolar giant planets (EGPs), and low-mass
stars.  The behavior of individual Fe-, Mg-, and Si-bearing gases and condensates is determined as a function of temperature, pressure, and metallicity.  Our results are thus independent of any
particular model atmosphere.  The condensation of Fe metal strongly affects iron chemistry by efficiently removing Fe-bearing species from the gas phase.  Monatomic Fe is the most abundant
Fe-bearing gas throughout the atmospheres of EGPs and L dwarfs and in the deep atmospheres of giant planets and T dwarfs. Mg- and Si-bearing gases are effectively removed from the atmosphere by
forsterite (Mg$_{2}$SiO$_{4}$) and enstatite (MgSiO$_{3}$) cloud formation. Monatomic Mg is the dominant magnesium gas throughout the atmospheres of EGPs and L dwarfs and in the deep atmospheres
of giant planets and T dwarfs. Silicon monoxide (SiO) is the most abundant Si-bearing gas in the deep atmospheres of brown dwarfs and EGPs, whereas SiH$_{4}$ is dominant in the deep atmosphere
of Jupiter and other gas giant planets. Several other Fe-, Mg-, and Si-bearing gases become increasingly important with decreasing effective temperature.   In principle, a number of Fe, Mg, and
Si gases are potential tracers of weather or diagnostic of temperature in substellar atmospheres.
\end{abstract}


\section{Introduction}\label{s Introduction}

Thermochemical models have been used to describe alkali \citep{lodders1999}, titanium and vanadium \citep{lodders2002apj}, carbon, nitrogen, and oxygen \citep{lodders2002}, sulfur and phosphorus
\citep{visscher2006}, condensate \citep[e.g.,][]{lodders2006,lodders2009exo} chemistry in the atmospheres of gas giant planets, brown dwarfs, and low-mass dwarf stars. Here we continue and
extend these previous studies by using thermochemical equilibrium calculations to model the chemical behavior of Fe, Mg, and Si in substellar objects.  Iron, magnesium, and silicon are the most
abundant rock-forming elements in a solar composition gas, and -- condensed as iron metal (Fe), and forsterite (Mg$_{2}$SiO$_{4}$) and enstatite (MgSiO$_{3}$) -- will produce the most massive
cloud layers in substellar atmospheres. Cloud formation strongly affects the optical and infrared spectra of substellar objects by removing gases from the overlying atmosphere and by introducing
solid or liquid cloud particles \citep[e.g.,][]{marley1996,lodders1999,burrows2000,lodders2006,visscher2006,lodders2009exo}.

The large and growing number of discovered brown dwarfs ($>$750) and extrasolar planets ($>$400) makes it impractical to model the thermochemistry of all objects individually
\citep{lodders2002}, as was done for Jupiter and Saturn \citep{fegley1994} and Gliese 229B \citep{fegley1996}.  Instead, we adopt an approach similar to that of our previous papers
\citep{lodders2002,visscher2006} and determine the abundance of each chemical species as a function of pressure, temperature, and metallicity, plotted in abundance contour diagrams. Our results
are thus independent of any particular pressure-temperature profile, and in principle, the atmospheric profile for any object may be superimposed on the abundance diagrams to determine its
equilibrium atmospheric chemistry. In some instances, the behavior of key gases may be diagnostic of atmospheric temperature and/or tracers of weather in substellar objects.

We begin with a brief description of our computational method ($\S$\ref{s Computational Method}), and then present our results for iron chemistry in substellar atmospheres in $\S$\ref{s Iron
Chemistry}. We first give an overview of iron chemistry in a solar composition gas and identify important gases and condensates ($\S$\ref{ss Overview of Iron Chemistry}).  This is followed by
more detailed discussion of the chemical behavior individual Fe-bearing gases as a function of temperature, pressure, and metallicity ($\S$\ref{ss Chemical Behavior of Iron-bearing Gases}).
Wherever possible, we note relevant spectroscopic observations of Fe-bearing gases in substellar objects.  Our results for iron are then summarized by illustrating Fe gas chemistry along the
atmospheric profiles of representative substellar objects ($\S$\ref{ss Iron Chemistry in Substellar Objects}). A similar approach to magnesium and silicon chemistry follows in $\S$\ref{s
Magnesium Chemistry} and $\S$\ref{s Silicon Chemistry}, respectively. We conclude with a brief summary in $\S$\ref{s Summary}.

\section{Computational Method}\label{s Computational Method}

Thermochemical equilibrium calculations were performed using a Gibbs free energy minimization code, previously used for modeling the atmospheric chemistry of Saturn \citep{visscher2006} and
sulfur and phosphorus chemistry in substellar objects \citep{visscher2006}. Thermodynamic data for the equilibrium calculations were taken from the compilations of \citet{gurvich1989},
\citet{robie1995}, the fourth edition of the JANAF Tables \citep{chase1998}, and the thermodynamic database maintained in the Planetary Chemistry Laboratory \citep{lodders2002}. This database
includes additional thermodynamic data from the literature for compounds absent from the other compilations.

All calculations were conducted using elemental abundances from \citet{lodders2003} for a solar system (i.e., protosolar) composition gas.  The effect of metallicity on sulfur and phosphorus
chemistry was examined by running computations at [Fe/H] = -0.5 dex (subsolar), [Fe/H] = 0 dex (solar), and [Fe/H] = +0.5 dex (enhanced) metallicities.  The metallicity factor, $m$, is defined
as $\log m = [\textrm{Fe/H}]$.  We assume that the elemental abundance ratios for Mg, Si, and other elements of interest vary similarly with [Fe/H] (e.g., [Mg/H] $\approx$ [Si/H] $\approx$
[Fe/H]) over the range of metallicities considered here \citep[see][]{edvardsson1993,chen2000,huang2005}.  When considering the chemical behavior of individual gases, we focus on higher
temperatures (800 K and higher), where thermochemical processes are expected to dominate over disequilibrium processes such as photochemistry or atmospheric mixing \citep[e.g.,
see][]{visscher2006}.

We assume that condensates settle gravitationally into a cloud layer and are removed from the cooler, overlying atmosphere. This equilibrium cloud condensate scenario for the deep atmospheres of
giant planets and brown dwarfs is supported by several lines of evidence \citep[e.g., see][and references therein]{lodders2006}.  First, the presence of germane (GeH$_{4}$) and the absence of
silane (SiH$_{4}$) in the upper atmospheres of Jupiter and Saturn (even though Si is expected to be much more abundant than Ge) can be explained by the removal of Si from the gas into silicate
clouds deeper in the atmosphere, whereas Ge remains in the gas phase \citep{fegley1988, fegley1994}. Secondly, the detection of H$_{2}$S in Jupiter's troposphere by the \textit{Galileo} entry
probe indicates that Fe must be sequestered into a cloud layer at deep atmospheric levels, because the formation of FeS would otherwise remove H$_{2}$S from the gas above the $\sim700$ K level
\citep{lodders2002,visscher2006}. Third, absorption from monatomic K gas in the spectra of T dwarfs \citep{burrows2000, geballe2001} requires the removal of Al and Si at deeper atmospheric
levels, because K would otherwise be removed from the observable atmosphere by the condensation of orthoclase (KAlSi$_{3}$O$_{8}$) \citep{lodders2006}. The presence of monatomic Na gas in brown
dwarfs \citep{kirkpatrick1999, burgasser2003apj, mclean2003, cushing2005} also suggests Al and Si removal, because albite (NaAlSi$_{3}$O$_{8}$) condensation would otherwise effectively remove Na
from the observable atmosphere. Furthermore, the removal of Na by Na$_{2}$S cloud formation is consistent with the observed weakening of Na atomic lines throughout the L dwarf spectral sequence
and their disappearance in early T dwarfs \citep[e.g.,][]{kirkpatrick1999,mclean2003,visscher2006}.  Finally, as we note below, the disappearance of iron, magnesium, and silicon spectral
features in later spectral types is consistent with removal of these elements into cloud layers. In our thermochemical model, the abundances of Fe-, Mg-, and Si-bearing gases above the clouds
are computed assuming saturation (equilibrium) vapor pressure. If supersaturation occurs, a condensate will form and settle toward the cloud layer to restore equilibrium.

\section{Iron Chemistry}\label{s Iron Chemistry}
\subsection{Overview of Iron Chemistry}\label{ss Overview of Iron Chemistry}

\begin{figure}
\scalebox{.7}{\includegraphics{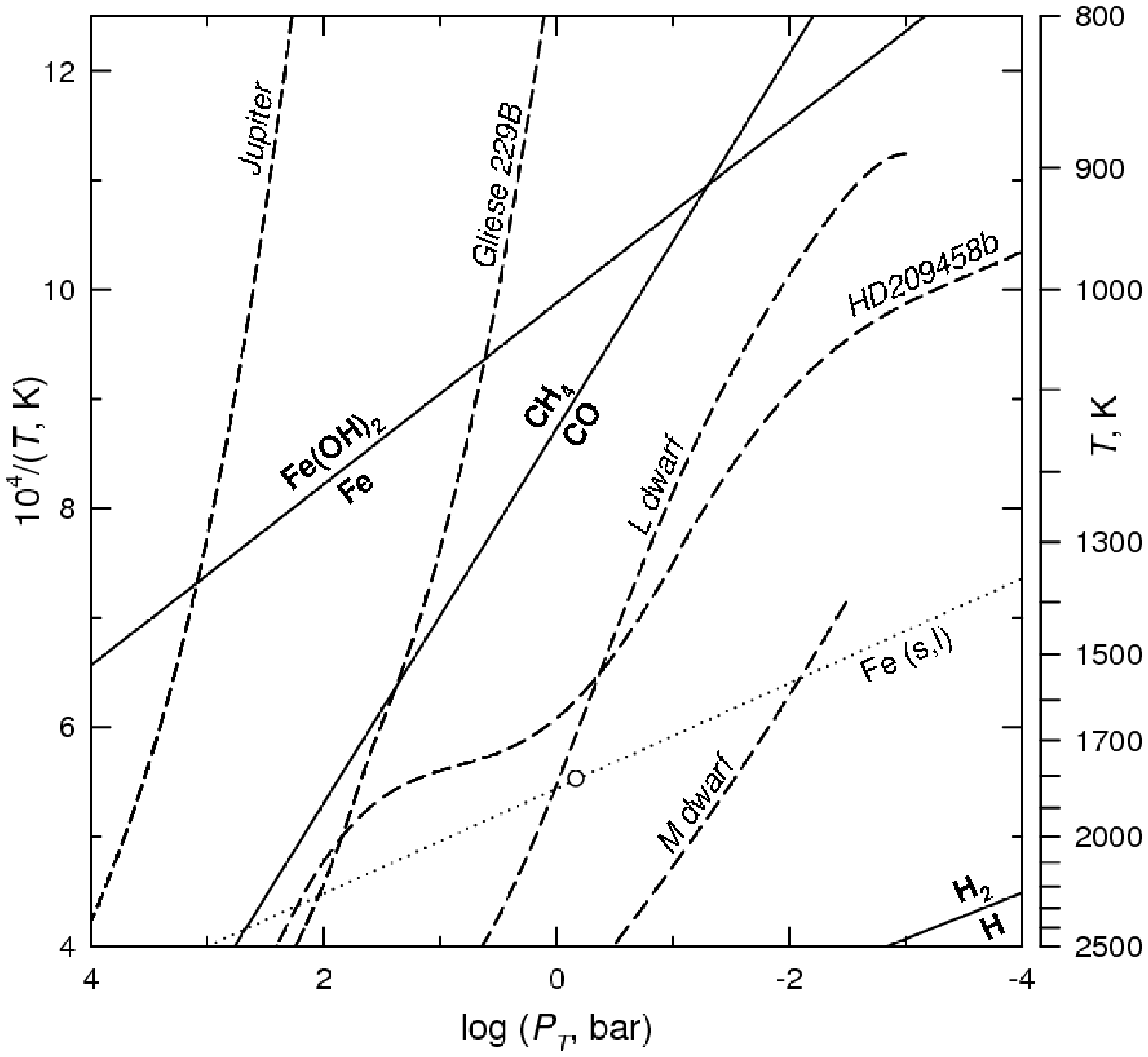}} \caption[Overview of iron chemistry]{Overview of iron chemistry as a function of temperature and pressure in a protosolar composition gas.  The solid line
indicates where Fe and Fe(OH)$_{2}$ have equal abundances.  The dotted line labeled Fe(s,l) is the condensation curve for iron, with an open circle denoting its normal melting point (1809 K).
Also shown are model atmospheric profiles for representative substellar objects (dashed lines). See text for details.} \label{figure iron chemistry overview}
\end{figure}

Figure \ref{figure iron chemistry overview} illustrates model atmospheric profiles for an M dwarf \citep[$T_{\textrm{\scriptsize{eff}}}$ = 2600 K, $\log g=5.0$;][]{tsuji1996}, an L dwarf
\citep[$T_{\textrm{\scriptsize{eff}}}$ = 1800 K, $\log g=5.0$;][]{burrows2006}, the hot, close-in (Pegasi) planet HD209458b \citep[$T_{\textrm{\scriptsize{eff}}}$ = 1350 K;][]{iro2005}, the T
dwarf Gliese 229B \citep[$T_{\textrm{\scriptsize{eff}}}$ = 960 K;][]{marley1996},  and Jupiter ($T_{\textrm{\scriptsize{eff}}}$ = 124 K), indicated by dashed lines. We note that Jovian
atmospheric chemistry differs slightly than that for a solar-metallicity gas because Jupiter has a heavy element enrichment comparable to [Fe/H] $\approx$ +0.5 dex \citep{lodders1999,
lodders2002}.

Also shown in Figure \ref{figure iron chemistry overview} are lines indicating where A(CH$_{4}$) = A(CO) and A(H$_{2}$) = A(H).  These boundaries are important because carbon and oxygen affect
the chemical behavior of many Fe, Mg, and Si-bearing gases.  Methane is the dominant carbon-bearing gas in Jupiter and T dwarfs (such as Gliese 229B) whereas CO is the dominant carbon-bearing
gas in L dwarfs and Pegasi planets (such as HD209458b).  Molecular hydrogen dissociates into monatomic H at high temperatures and low pressures (lower right corner of Figure \ref{figure iron
chemistry overview}). However, H$_{2}$ is the dominant form of hydrogen in substellar objects, and we therefore take $X_{\textrm{\scriptsize{H}}_{2}}\approx
X_{\Sigma\textrm{\scriptsize{H}}}\approx 0.84$ throughout the following.  This approximation holds for metallicities up to [Fe/H] $\approx$ +0.5 dex; at higher metallicities the H$_{2}$ mole
fraction abundance decreases as the relative abundance of heavy elements increases (e.g., $X_{\textrm{\scriptsize{H}}_{2}}\approx0.82$ at [Fe/H] $\approx$ +1.0 dex).

The dotted line in Figure \ref{figure iron chemistry overview} shows the condensation curve for Fe metal, with an open circle denoting its normal melting point (1809 K). Iron condensation occurs
via
\begin{equation}\label{reaction Fe condensation}
\textrm{Fe}=\textrm{Fe(s,l)}.
\end{equation}
The equilibrium condensation temperature ($T_{{\textrm{\scriptsize{cond}}}}$) of Fe is approximated by
\begin{equation}\label{condensation curve Fe}
10^{4}/T_{\textrm{\scriptsize{cond}}}(\textrm{Fe} )= 5.44 - 0.48\log P_{T} - 0.48[\textrm{Fe/H}],
\end{equation}
where higher pressures and/or metallicities lead to higher Fe condensation temperatures.  Iron cloud formation effectively removes nearly all
iron from the atmosphere, and the abundances of Fe-bearing gases above the clouds rapidly decrease with altitude.

Monatomic Fe gas is the dominant Fe-bearing gas in a protosolar composition gas until it is replaced by Fe(OH)$_{2}$ at low temperatures and
high pressures. The conversion between Fe and Fe(OH)$_{2}$ is represented by the net thermochemical reaction
\begin{equation}\label{reaction Fe:Fe(OH)2}
\textrm{Fe}+2\textrm{H}_{2}\textrm{O}=\textrm{Fe(OH)}_{2}+\textrm{H}_{2},
\end{equation}
and the solid line dividing the Fe and Fe(OH)$_{2}$ fields in Figure \ref{figure iron chemistry overview} indicates where these gases have equal
abundances A(Fe) = A(Fe(OH)$_{2}$) $\approx \frac{1}{2}\Sigma$Fe$_{\textrm{\scriptsize{gas}}}$, where $\Sigma$Fe$_{\textrm{\scriptsize{gas}}}$
is the total amount of iron in the gas.  The position of this line is given by
\begin{equation}\label{boundary line Fe:Fe(OH)2}
\log P_{T} = 11.94 - 12088/T - 2[\textrm{Fe/H}],
\end{equation}
showing that an increase in metallicity will shift the Fe-Fe(OH)$_{2}$ boundary to higher temperatures and lower pressures.  In other words, an increase in metallicity increases the stability
field of Fe(OH)$_{2}$ for otherwise constant conditions.

\subsection{Chemical Behavior of Iron-bearing Gases}\label{ss Chemical Behavior of Iron-bearing Gases}
\subsubsection{Monatomic Iron, Fe}\label{sss Monatomic Iron, Fe}
The mole fraction abundance of monatomic Fe as a function of pressure and temperature is shown in Figure \ref{figure iron gas 1}a.  As mentioned
above, monatomic Fe is the dominant Fe-bearing gas over a wide range of pressures and temperatures in a solar system composition gas.  Below the
Fe clouds, the abundance of Fe gas is representative of the total iron content of the atmosphere ($X_{\textrm{\scriptsize{Fe}}}\approx
X_{\Sigma\textrm{\scriptsize{Fe}}}$), and its abundance is given by
\begin{equation}\label{equation Fe below clouds}
\log X_{\textrm{\scriptsize{Fe}}}\approx -4.24 + [\textrm{Fe/H}].
\end{equation}
Upon Fe metal condensation, the amount of iron in the gas rapidly decreases and the equilibrium Fe gas abundance is governed by its vapor
pressure over solid or liquid iron, represented in reaction (\ref{reaction Fe condensation}).  The mole fraction abundance of monatomic Fe above
the clouds is given by
\begin{equation}\label{equation Fe above clouds}
\log X_{\textrm{\scriptsize{Fe}}}\approx 7.23 - 20995/T - \log P_{T},
\end{equation}
inversely proportional to $P_{T}$.  The Fe gas abundance is independent of metallicity in this region because it depends solely on the temperature-dependent vapor pressure of iron.  As described
below (see $\S$\ref{sss Iron Hydride, FeH} and $\S$\ref{sss Iron Monohydroxide, FeOH}), this expression can be used with chemical equilibria to determine the abundances of other iron gases in
substellar atmospheres. Equations giving the abundance of Fe (and other Fe-bearing gases) as a function of temperature, pressure, and metallicity below and above the Fe clouds are listed in
Table \ref{table iron reactions}.

Neutral monatomic Fe possesses several hundred spectral lines in the $J$ and $H$ bands \citep{melendez1999}, and Fe I features are observed in the spectra of brown dwarfs and low-mass dwarf
stars \citep[e.g.][]{cushing2005}. The strong Fe feature at 1.189 $\mu$m weakens in mid- to late-type M dwarf spectra and generally disappears in mid-type L dwarfs \citep{mclean2003,
cushing2005}. This trend is consistent with the removal of iron from the gas into an Fe metal cloud deck located deeper below the photosphere in objects with low effective temperatures
\citep[e.g.][]{burgasser2002apjl}.

\subsubsection{Iron Hydride, FeH}\label{sss Iron Hydride, FeH} 
\begin{figure}
\scalebox{.7}{\includegraphics{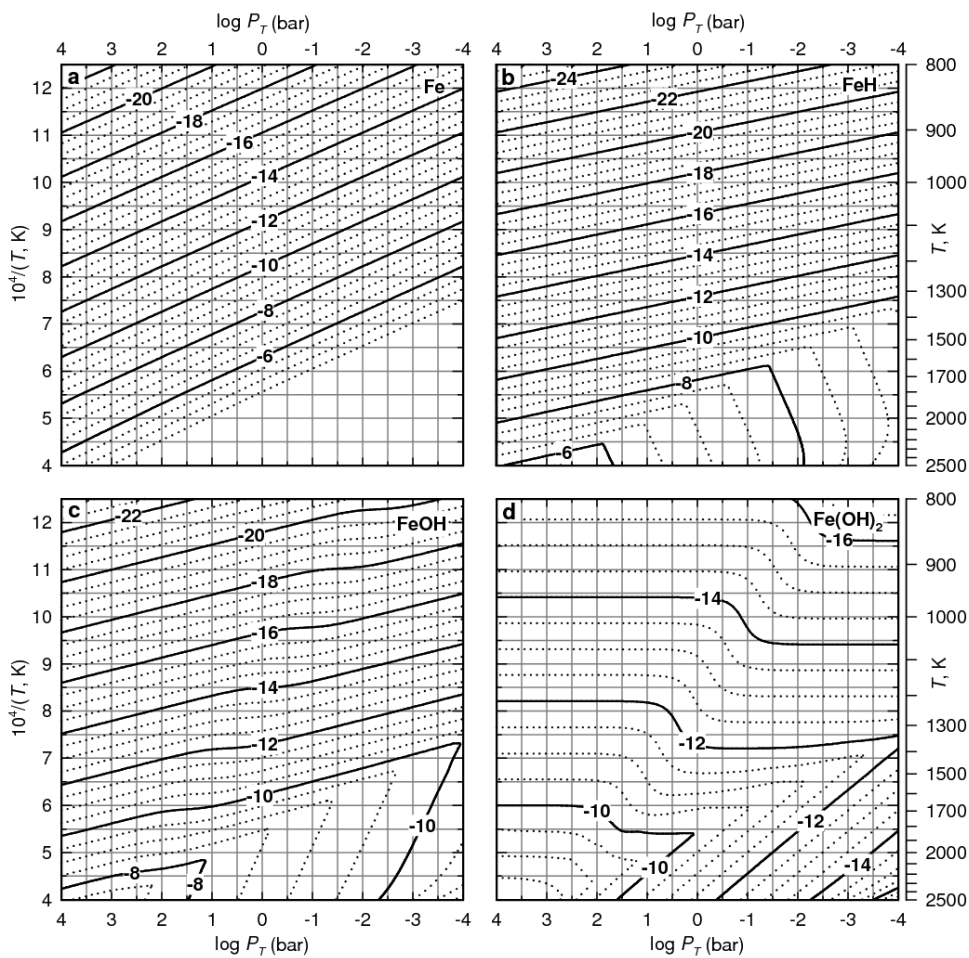}} \caption[Individual Fe-bearing gases 1]{Mole fraction contours (on a logarithmic scale) for (a) monatomic iron (Fe), (b) iron hydride (FeH), (c) iron
monohydroxide (FeOH), and (d) iron hydroxide (Fe(OH)$_{2}$) as a function of pressure and temperature in a solar-metallicity gas.  The kinks in the abundance contours are due to Fe metal or
liquid condensation.} \label{figure iron gas 1}
\end{figure}

The chemical behavior of FeH in a protosolar composition gas is illustrated in
Figure \ref{figure iron gas 1}b.  The conversion between Fe and FeH occurs via the net thermochemical reaction
\begin{equation}\label{reaction Fe:FeH}
\textrm{Fe} + 0.5\textrm{H}_{2} = \textrm{FeH}
\end{equation}
Here we show how chemical equilibria may used to derive equations giving the abundance of Fe-bearing species as a function of temperature,
pressure, and metallicity. Rearranging the equilibrium constant expression for reaction (\ref{reaction Fe:FeH}) yields
\begin{equation}\label{equation FeH abundance generic}
\log X_{\textrm{\scriptsize{FeH}}} = \log X_{\textrm{\scriptsize{Fe}}} + 0.5\log X_{\textrm{\scriptsize{H}}_{2}} + 0.5\log P_{T} + \log
K_{\ref{reaction Fe:FeH}}.
\end{equation}
Using the Fe abundance from equation (\ref{equation Fe below clouds}), the temperature dependence of $K_{\ref{reaction Fe:FeH}}$ ($\log K_{\ref{reaction Fe:FeH}} \approx -1.85 - 1905/T$ from 800
to 2500 K), and the hydrogen abundance ($X_{\textrm{\scriptsize{H}}_{2}}\approx0.84$), the FeH abundance between the H$_{2}$-H boundary and the Fe cloud deck is given by
\begin{equation}\label{equation FeH abundance below clouds}
\log X_{\textrm{\scriptsize{FeH}}} \approx -6.13 - 1905/T + 0.5 \log P_{T} + [\textrm{Fe/H}],
\end{equation}
proportional to $P_{T}^{0.5}$ and $m$.  At high temperatures ($>2000$ K) and low pressures ($<10^{-2}$ bar) as monatomic H becomes increasingly abundant the H$_{2}$ abundance begins to decrease
near the H$_{2}$-H boundary. The reduced H$_{2}$ abundance, in turn, reduces FeH formation via reaction (\ref{reaction Fe:FeH}) and changes the chemical behavior of FeH at high temperatures and
low pressures (i.e., lower right corner of Figure \ref{figure iron gas 1}a).

Above the iron clouds, the Fe abundance in equation (\ref{equation Fe above clouds}) is used in equation (\ref{equation FeH abundance generic}) to give
\begin{equation}\label{equation FeH abundance above clouds}
\log X_{\textrm{\scriptsize{FeH}}} \approx 5.34 - 22900/T - 0.5 \log P_{T},
\end{equation}
where $X_{\textrm{\scriptsize{FeH}}}$ is proportional to $P_{T}^{-0.5}$ and independent of metallicity.  The differences in chemical behavior of FeH below (equation \ref{equation FeH abundance
below clouds}) and above (equation \ref{equation FeH abundance above clouds}) the Fe cloud deck are illustrated in the shape of the FeH abundance contours in Figure \ref{figure iron gas 1}b. The
sharp bends in the contours correspond to the Fe condensation curve in Figure \ref{figure iron chemistry overview}.

Absorption bands from FeH are common in the near-infrared spectra of brown dwarfs \citep[e.g.][]{kirkpatrick1999, mclean2003, cushing2003,cushing2005}. The most prominent FeH feature is the band
located at 0.9896 $\mu$m, which weakens throughout the L dwarf spectral sequence and in early T dwarfs \citep{mclean2003, nakajima2004, cushing2005}, consistent with the removal of iron into a
cloud located deeper and deeper in the atmosphere.  This feature unexpectedly strengthens again in mid-type T dwarf spectra \citep{burgasser2002apj,burgasser2002apjl, nakajima2004, cushing2005},
prompting different explanations for the shape of the color-magnitude diagram for brown dwarfs near the L-T transition.  \citet{nakajima2004} suggested that the strengthening FeH bands are
caused by upward convective mixing of FeH gas {from} deeper levels where it is more abundant.  However, as pointed out by \citet{burgasser2002apjl} and demonstrated \citet{lodders2006}, the
fragile Fe$\sbond$H bond is unlikely to survive convective upwelling in a T dwarf atmosphere. Instead, the observations are plausibly explained by cloud disruption and clearing (in a $\sim$1
$\mu$m window) which allows the observation of FeH gas {at} deep atmospheric levels \citep{burgasser2002apjl,lodders2006}.

\subsubsection{Iron Monohydroxide, FeOH}\label{sss Iron Monohydroxide, FeOH}
Mole fraction contours of FeOH are illustrated in Figure \ref{figure iron gas 1}c.  The equilibrium between Fe and FeOH is represented by the
net thermochemical reaction
\begin{equation}\label{reaction Fe:FeOH}
\textrm{Fe}+\textrm{H}_{2}\textrm{O}=\textrm{FeOH}+0.5\textrm{H}_{2},
\end{equation}
and expressions giving the FeOH abundance as a function of temperature, pressure, and metallicity are listed in Table \ref{table iron
reactions}. Above the iron clouds, the curvature in the FeOH abundance contours along the CH$_{4}$-CO boundary results from the effect of carbon
chemistry on the H$_{2}$O abundance in reaction (\ref{reaction Fe:FeOH}) \citep[e.g.][]{lodders2002}.  For example, the atmospheric water
abundance may be written as
\begin{equation}\label{equation H2O abundance generic}
\log X_{\textrm{\scriptsize{H}}_{2}\textrm{\scriptsize{O}}} = \log X_{\textrm{\scriptsize{H}}_{2}\textrm{\scriptsize{O}}}'+[\textrm{Fe/H}],
\end{equation}
where $X_{\textrm{\scriptsize{H}}_{2}\textrm{\scriptsize{O}}}'$ is the water abundance in a solar-metallicity gas, which is
$\textrm{log}_{10}X_{\textrm{\scriptsize{H}}_{2}\textrm{\scriptsize{O}}}'\approx10^{-3.12}$ inside the CH$_{4}$ field and
$X_{\textrm{\scriptsize{H}}_{2}\textrm{\scriptsize{O}}}'\approx10^{-3.58}$ inside the CO field in Figure \ref{figure iron chemistry overview}. At temperatures and pressures near the CH$_{4}$-CO
boundary, $X_{\textrm{\scriptsize{H}}_{2}\textrm{\scriptsize{O}}}'$ may be derived from CH$_{4}$-CO equilibria \citep[][]{lodders2002}. Rearranging the equilibrium constant expression for
reaction (\ref{reaction Fe:FeOH}) gives
\begin{equation}\label{equation FeOH abundance generic}
\log X_{\textrm{\scriptsize{FeOH}}} = \log X_{\textrm{\scriptsize{Fe}}} + \log X_{\textrm{\scriptsize{H}}_{2}\textrm{\scriptsize{O}}}-0.5\log
X_{\textrm{\scriptsize{H}}_{2}}+0.5 \log P_{T}+\log K_{\ref{reaction Fe:FeOH}}.
\end{equation}
 Substituting for the Fe abundance from equation (\ref{equation Fe above clouds}), the H$_{2}$O abundance from equation (\ref{equation H2O abundance generic}), the H$_{2}$ abundance
($X_{\textrm{\scriptsize{H}}_{2}}\approx 0.8384$), and the temperature dependence of $K_{\ref{reaction Fe:FeOH}}$ ($\log K_{\ref{reaction
Fe:FeOH}}\approx -2.25+2468/T$ from 800 to 2500 K), equation (\ref{equation FeOH abundance generic}) becomes
\begin{equation}\label{equation FeOH abundance above clouds}
\log X_{\textrm{\scriptsize{FeOH}}} \approx 5.02 - 18527/T - 0.5 \log P_{T} + 2[\textrm{Fe/H}] + \log X_{\textrm{\scriptsize{H}}_{2}\textrm{\scriptsize{O}}}'.
\end{equation}
This expression gives the FeOH abundance above the iron clouds and includes the effect of metallicity on the atmospheric water abundance.  The kinks in the FeOH contours denote the position of
the Fe condensation curve.  As demonstrated in equation (\ref{equation FeOH abundance above clouds}) and shown in Figure \ref{figure iron gas 1}c, the mole fraction abundance of FeOH is
proportional to $P_{T}^{-0.5}$ throughout this region.

\subsubsection{Iron Hydroxide, Fe(OH)$_{2}$}\label{sss Iron Hydroxide, Fe(OH)2}
The chemical behavior of Fe(OH)$_{2}$ as a function of temperature and pressure is illustrated in Figure \ref{figure iron gas 1}d.  The equilibrium conversion between Fe and Fe(OH)$_{2}$ is
represented by reaction (\ref{reaction Fe:Fe(OH)2}).  Below the Fe clouds (kinks in abundance contours), the Fe(OH)$_{2}$ abundance increases with total pressure, and has a strong $m^3$
dependence on metallicity (see Table \ref{table iron reactions}).

The Fe(OH)$_{2}$ abundance in reaction (\ref{reaction Fe:Fe(OH)2}) is very sensitive to the H$_{2}$O abundance since the formation of one mole of iron hydroxide requires two moles of water.  As
a result, there is a large shift in the Fe(OH)$_{2}$ abundance contours above the clouds (inflections in Figure \ref{figure iron gas 1}d) when moving between the CH$_{4}$ and CO fields.  Within
each field, the Fe(OH)$_{2}$ abundance above the clouds is essentially pressure-independent (see Table\ref{table iron reactions}) and is therefore, in principle, diagnostic of atmospheric
temperature.

As shown in Figure \ref{figure iron chemistry overview}, Fe(OH)$_{2}$ becomes the most abundant Fe-bearing gas at low temperatures and high pressures in a solar composition gas.  Iron hydroxide
is therefore expected to be the most abundant iron gas at $T\la1600$ on Jupiter ($[\textrm{Fe/H}]\approx+0.5$) and $T\la 1070$ on Gliese 229B.  However, we emphasize that the amount of iron
remaining in the gas is greatly diminished by the condensation of Fe metal at deeper atmospheric levels. For example, even where it is the most abundant Fe-bearing gas, the predicted
Fe(OH)$_{2}$ mole fraction is $\approx10^{-12.6}$ at the 1000 K level in Jupiter's atmosphere.

\subsubsection{Iron Monoxide, FeO}\label{sss Iron Monoxide, FeO}
\begin{figure}
\scalebox{.7}{\includegraphics{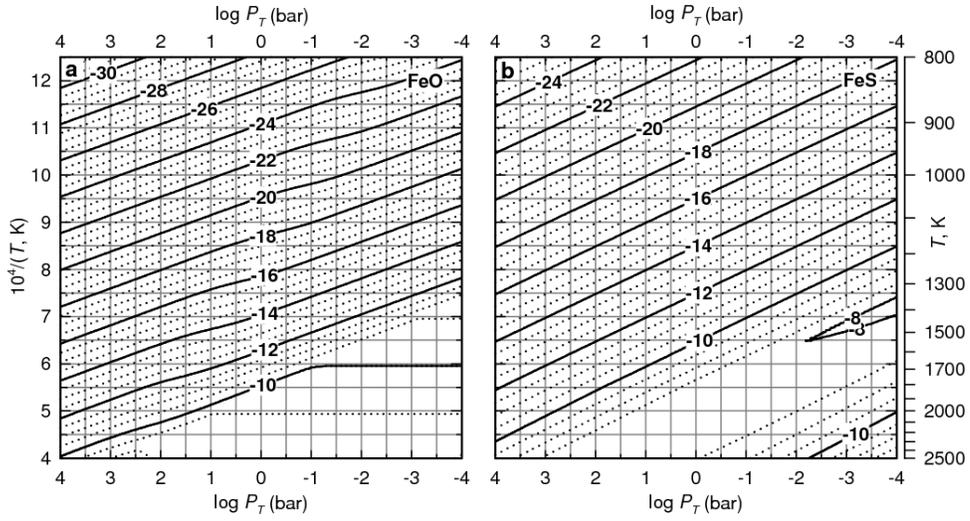}} \caption[Individual Fe-bearing gases 2]{Mole fraction contours (on a logarithmic scale) for (a) iron monoxide (FeO) and (b) iron sulfide (FeS) as a
function of pressure and temperature in a solar-metallicity gas.} \label{figure iron gas 2}
\end{figure}

Mole fraction abundance contours for FeO are shown in Figure \ref{figure iron gas 2}a for a solar composition gas.  Monatomic iron reacts with water to form FeO via the net reaction
\begin{equation}\label{reaction Fe:FeO}
\textrm{Fe}+\textrm{H}_{2}\textrm{O}=\textrm{FeO}+\textrm{H}_{2}.
\end{equation}
Between the H$_{2}$-H boundary and the Fe cloud deck, $X_{\textrm{\scriptsize{FeO}}}$ is proportional to $m^{2}$ and is independent of pressure (see Table \ref{table iron reactions}).  In
principle, the FeO abundance in this region is therefore diagnostic of either temperature or metallicity. Above the clouds, the FeO abundance decreases with pressure and increases with
metallicity. As for FeOH and Fe(OH)$_{2}$, the FeO abundance is sensitive to the H$_{2}$O abundance and the FeO abundance contours display a subtle shift along the CH$_{4}$-CO equal abundance
boundary.

\subsubsection{Iron Sulfide, FeS}\label{sss Iron Sulfide, FeS}
The abundance of FeS gas as a function of temperature and pressure is illustrated in Figure \ref{figure iron gas 2}b.  The chemical behavior of
FeS is governed by the reaction between Fe and H$_{2}$S:
\begin{equation}\label{reaction Fe:FeS}
\textrm{Fe}+\textrm{H}_{2}\textrm{S}=\textrm{FeS}+\textrm{H}_{2}.
\end{equation}
Above the iron clouds, H$_{2}$S is the dominant sulfur-bearing gas ($\log X_{\textrm{\scriptsize{H}}_{2}\textrm{\scriptsize{S}}}\approx-4.52+[\textrm{Fe/H}]$), and the FeS abundance is
proportional to $P_{T}^{-1}$ and $m$ (see Table \ref{table iron reactions}).  Below the Fe cloud deck, the FeS abundance is $\sim10$ ppb and decreases at higher temperatures and lower pressures
as H$_{2}$S is replaced by SH and S \citep{visscher2006}.  Iron sulfide is predicted to be the second or third most abundant iron gas throughout L dwarf atmospheres and in the upper atmospheres
($< 1500$ K) of Pegasi planets.

\subsection{Iron Chemistry in Substellar Objects}\label{ss Iron Chemistry in Substellar Objects}
\begin{figure}
\scalebox{.7}{\includegraphics{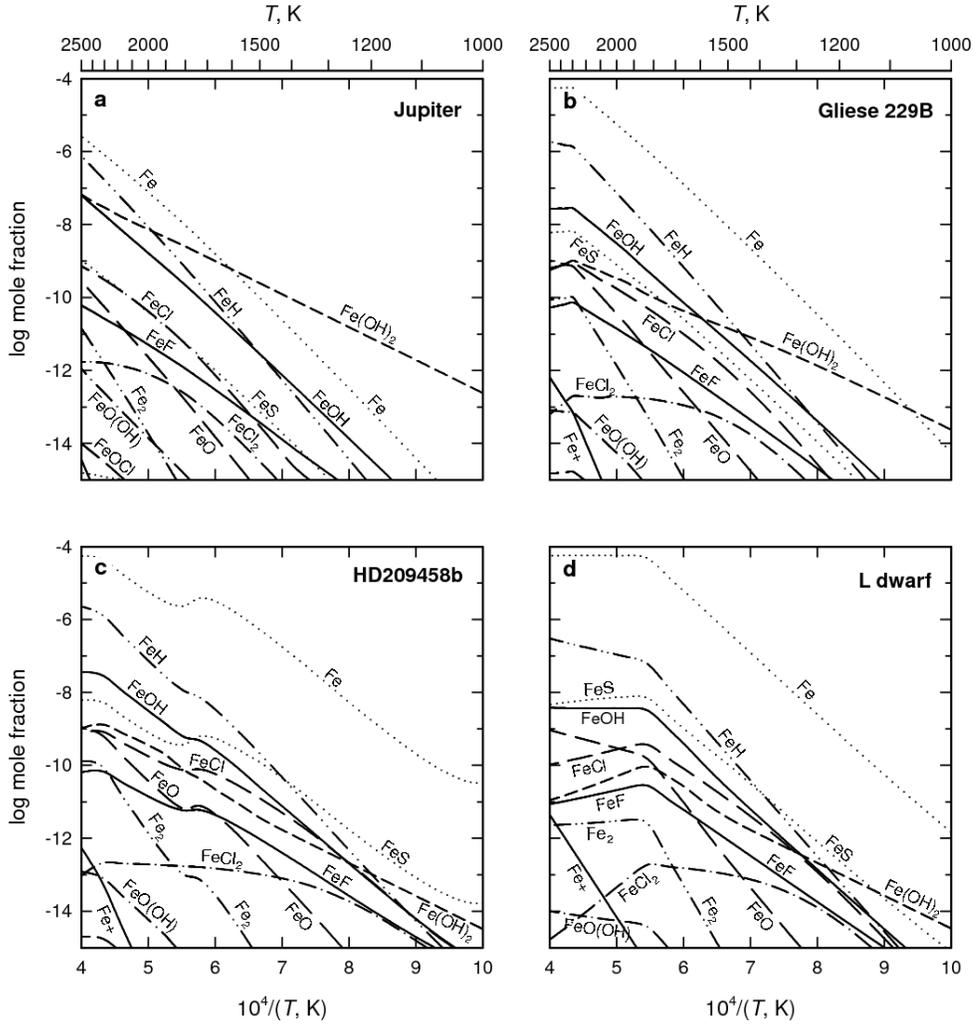}} \caption[Summary of iron chemistry]{Iron chemistry along the atmospheric pressure-temperature profile from 1000 to 2500 K for (a) Jupiter, (b) Gliese
229B, (c) HD209458b, and (d) an L dwarf ($T_{\textrm{\scriptsize{eff}}}=1800$ K).  Breaks in the Fe curves show where Fe condenses (b,d).  Iron is condensed throughout the entire temperature
range shown here for Jupiter and HD209458b (a,c).} \label{figure iron chemistry summary}
\end{figure}

Figures \ref{figure iron chemistry summary}a-\ref{figure iron chemistry summary}d summarize the iron gas chemistry in four representative substellar objects: Jupiter, the T dwarf Gliese 229B,
the Pegasi planet HD209458b, and an L dwarf ($T_{\textrm{\scriptsize{eff}}}=1800$ K). Iron chemistry in substellar objects is strongly affected by Fe metal condensation at deep atmospheric
levels.  Monatomic Fe is the dominant Fe-bearing gas in the deep atmospheres of Jupiter and Gliese 229B, and throughout the atmospheres of HD209458b and the L dwarf.  A number of Fe-bearing
gases become relatively more abundant in objects with lower effective temperatures. On Jupiter and Gliese 229B, Fe(OH)$_{2}$ replaces Fe at lower temperatures.  The Fe(OH)$_{2}$ and FeO
abundances are pressure-independent and thus potentially diagnostic of atmospheric temperature, respectively, above and below the iron clouds. For all four objects, iron hydride (FeH) is the
second most abundant Fe-bearing gas at deep atmospheric levels, until it is replaced at lower temperatures by Fe(OH)$_{2}$, FeOH, or FeS. Because of its strong absorption and relatively high
abundance near the iron condensation level, FeH is a tracer of weather in brown dwarfs \citep[e.g.][]{burgasser2002apjl}. Other Fe-bearing gases (e.g. Fe, FeOH, FeS) are potential tracers of
weather since they typically achieve their maximum abundance near iron cloud base.

\section{Magnesium Chemistry}\label{s Magnesium Chemistry}
\subsection{Overview of Magnesium Chemistry}\label{ss Overview of Magnesium Chemistry}
\begin{figure}
\scalebox{.7}{\includegraphics{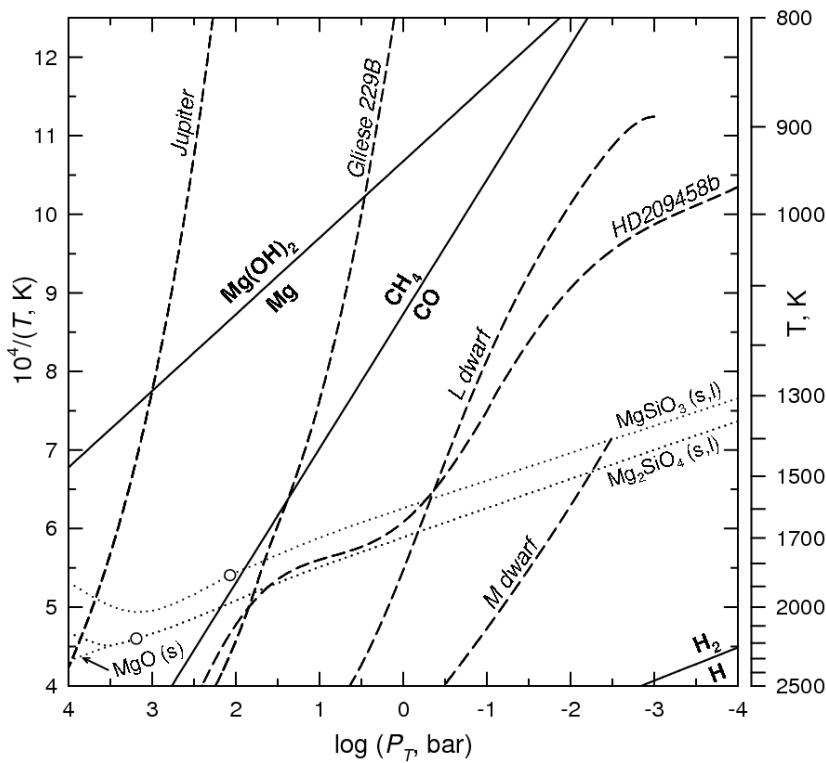}} \caption[Overview of magnesium chemistry]{Overview of magnesium chemistry as a function of temperature and pressure in a protosolar composition gas.  The
solid line indicates where Mg and Mg(OH)$_{2}$ have equal abundances. The dotted lines labeled Mg$_{2}$SiO$_{4}$(s,l) and MgSiO$_{3}$(s,l) show the condensation temperatures of forsterite and
enstatite, with circles denoting their normal melting points of 1803 K (enstatite) and 2163 K (forsterite). The dotted line labeled MgO(s) is the condensation curve for periclase. See text for
details.} \label{figure magnesium chemistry overview}
\end{figure}

Figure \ref{figure magnesium chemistry overview} gives an overview of Mg chemistry as a function of pressure and temperature in a protosolar composition gas.  The dotted lines labeled
Mg$_{2}$SiO$_{4}$(s,l), MgSiO$_{3}$(s,l), and MgO(s) show the equilibrium condensation curves for forsterite, enstatite, and periclase, respectively, and the open circles denote the normal
melting temperatures for forsterite (2163 K) and enstatite (1830 K).  Forsterite (Mg$_{2}$SiO$_{4}$) condenses via the net thermochemical reaction
\begin{equation}\label{reaction forsterite condensation}
2\textrm{Mg}+3\textrm{H}_{2}\textrm{O}+\textrm{SiO}=\textrm{Mg}_{2}\textrm{SiO}_{4}\textrm{(s,l)}+3\textrm{H}_{2},
\end{equation}
and its condensation temperature as a function of pressure and metallicity is approximated by
\begin{equation}\label{condensation curve forsterite}
10^{4}/T_{\textrm{\scriptsize{cond}}}(\textrm{Mg}_{2}\textrm{SiO}_{4})\approx 5.89 - 0.37\log P_{T} - 0.73[\textrm{Fe/H}],
\end{equation}
At slightly lower temperatures, enstatite (MgSiO$_{3}$) condensation occurs via the net reaction
\begin{equation}\label{reaction enstatite condensation}
\textrm{Mg}+2\textrm{H}_{2}\textrm{O}+\textrm{SiO}=\textrm{MgSiO}_{3}\textrm{(s,l)}+2\textrm{H}_{2}.
\end{equation}
The enstatite condensation curve is approximated by
\begin{equation}\label{condensation curve enstatite}
10^{4}/T_{\textrm{\scriptsize{cond}}}(\textrm{MgSiO}_{3})\approx 6.26 - 0.35\log P_{T} - 0.70[\textrm{Fe/H}],
\end{equation}
where higher pressures and/or metallicities lead to higher condensation temperatures.  At very high pressures, forsterite and enstatite
condensation temperature are depressed as SiO is replaced by SiH$_{4}$ (see $\S$\ref{ss Overview of Silicon Chemistry}). Periclase (MgO)
condenses via the net thermochemical reaction
\begin{equation}\label{reaction periclase condensation}
\textrm{Mg}+\textrm{H}_{2}\textrm{O}=\textrm{MgO(s)}+\textrm{H}_{2},
\end{equation}
at pressures greater than $10^{3.5}$ bar.  The condensation curve for akermanite (Ca$_{2}$MgSi$_{2}$O$_{7}$) is not shown here because most Ca is expected to be removed at deeper atmospheric
levels by the condensation of refractory calcium aluminates \citep[e.g.,][]{lodders2002apj}.  If no calcium is removed, Ca$_{2}$MgSi$_{2}$O$_{7}$ condensation would consume $\sim3\%$ of the
total atmospheric Mg inventory.  In the same way, olivine ((Mg,Fe)$_{2}$SiO$_{4}$) and fayalite (Fe$_{2}$SiO$_{4}$) are not expected in substellar atmospheres because nearly all Fe is removed
from the gas phase by iron metal condensation at higher temperatures \citep[e.g.,][]{cushing2006,lodders2006}

Magnesium-silicate cloud formation is very effective at removing nearly all ($>$99\%) magnesium from the atmosphere, and the abundances of Mg-bearing gases rapidly decrease with altitude above
the clouds.  This behavior is generally consistent with the disappearance of Mg spectral features by early-type L dwarfs (see \S\ref{sss Monatomic Magnesium, Mg}).  Furthermore,
\citet{cushing2006} find a Si$\sbond$O absorption feature at 10 $\mu$m in mid-type L dwarfs which is consistent with the presence of silicate grains and a weak 9.17 $\mu$m feature tentatively
attributed to crystalline enstatite.

Monatomic Mg is the dominant Mg-bearing gas in substellar atmospheres until it is replaced by Mg(OH)$_{2}$ at low temperatures and high pressures. The conversion between Mg and Mg(OH)$_{2}$ is
represented by the net thermochemical reaction
\begin{equation}\label{reaction Mg:Mg(OH)2}
\textrm{Mg}+2\textrm{H}_{2}\textrm{O}=\textrm{Mg(OH)}_{2}+\textrm{H}_{2},
\end{equation}
and the solid line in Figure \ref{figure magnesium chemistry overview} indicates where Mg and Mg(OH)$_{2}$ have equal abundances A(Mg) =
A(Mg(OH)$_{2}$) $\approx \frac{1}{2}\Sigma$Mg$_{\textrm{\scriptsize{gas}}}$, where $\Sigma$Mg$_{\textrm{\scriptsize{gas}}}$ is the total amount
of magnesium in the gas.  The position of the equal-abundance line is given by
\begin{equation}\label{boundary line Mg:Mg(OH)2}
\log P_{T} = 10.96 - 10267/T - 2[\textrm{Fe/H}],
\end{equation}
where an increase in metallicity shifts the Mg-Mg(OH)$_{2}$ boundary to higher temperatures and lower pressures.

\subsection{Chemical Behavior of Magnesium-bearing Gases}\label{ss Chemical Behavior of Magnesium-bearing Gases}
\subsubsection{Monatomic Magnesium, Mg}\label{sss Monatomic Magnesium, Mg}
\begin{figure}
\scalebox{.7}{\includegraphics{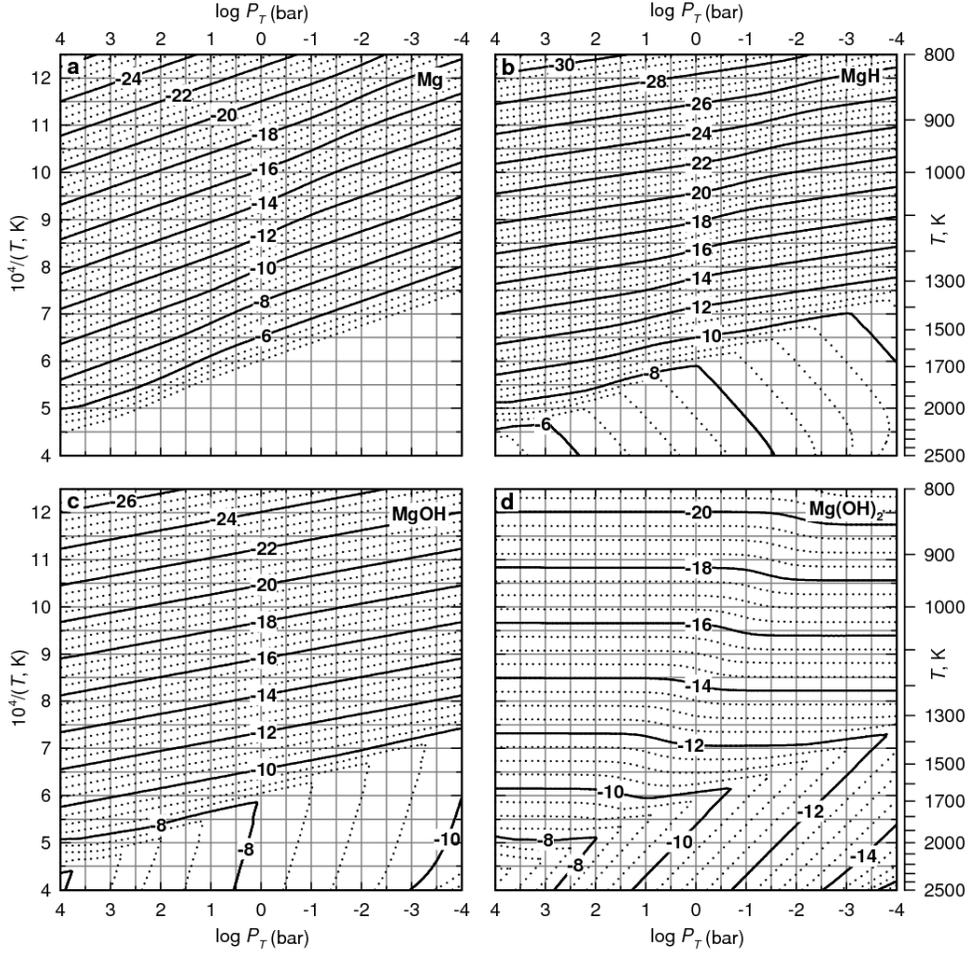}} \caption[Individual Mg-bearing gases 1]{Mole fraction contours (on a logarithmic scale) for (a) monatomic magnesium (Mg), (b) magnesium hydride (MgH), (c)
magnesium monohydroxide (MgOH), and (d) magnesium hydroxide (Mg(OH)$_{2}$) as a function of pressure and temperature in a solar-metallicity gas.  The kinks in the abundance contours occur at the
condensation temperature of Mg$_{2}$SiO$_{4}$ (forsterite).} \label{figure magnesium gas 1}
\end{figure}

Figure \ref{figure magnesium gas 1}a shows the chemical behavior of monatomic Mg as a function of pressure and temperature in a solar-metallicity gas.  The chemical behavior of Mg is strongly
affected by silicate cloud formation.  Below the forsterite clouds, the abundance of Mg gas is given by
\begin{equation}\label{equation Mg below clouds}
\log X_{\textrm{\scriptsize{Mg}}} \approx -4.15 + [\textrm{Fe/H}],
\end{equation}
and comprises nearly 100\% of the total elemental Mg content in the atmosphere ($X_{\textrm{\scriptsize{Mg}}}\approx X_{\Sigma\textrm{\scriptsize{Mg}}}$). Upon forsterite and enstatite
condensation, the magnesium gas abundance is governed by its vapor pressure over rock, represented by reactions (\ref{reaction forsterite condensation}) and (\ref{reaction enstatite
condensation}). Curvature in the Mg mole fraction contours in Figure \ref{figure magnesium gas 1}a occurs along the CH$_{4}$-CO boundary, which affects the H$_{2}$O abundance in reactions
(\ref{reaction forsterite condensation}) and (\ref{reaction enstatite condensation}). The Mg abundance is therefore
\begin{equation}\label{equation Mg above clouds CH4}
\log X_{\textrm{\scriptsize{Mg}}} \approx 11.37 -27250/T-\log P_{T} - [\textrm{Fe/H}].
\end{equation}
in CH$_{4}$-dominated objects, and
\begin{equation}\label{equation Mg above clouds CO}
\log X_{\textrm{\scriptsize{Mg}}} \approx 11.83 -27250/T-\log P_{T} - [\textrm{Fe/H}].
\end{equation}
in CO-dominated objects.  We can use these expressions along with chemical equilibria to determine the abundances of other Mg-bearing gases in
substellar atmospheres.  Equations giving the abundance of Mg-bearing gases as a function of pressure, temperature, and metallicity in a
protosolar composition gas are listed in Table \ref{table magnesium reactions}.

Several Mg absorption lines are present in the near infrared spectra of M dwarfs, including the prominent feature at 1.183 $\mu$m \citep{jones1996, mclean2003, cushing2005}.  These features
weaken in mid- to late-type M dwarfs and generally disappear by $\sim$ L1 \citep{cushing2005}, consistent with the removal of magnesium into Mg$_{2}$SiO$_{4}$ and MgSiO$_{3}$ clouds located at
increasingly greater depths below the observable atmosphere \citep[e.g., see][]{lodders2004science,lodders2006}.

\subsubsection{Magnesium Hydride, MgH}\label{sss Magnesium Hydride, MgH}
Mole fraction contours for MgH are illustrated in Figure \ref{figure magnesium gas 1}b.  The MgH abundance is governed by equilibrium with
monatomic Mg, via the net thermochemical reaction
\begin{equation}\label{reaction Mg:MgH}
\textrm{Mg}+0.5\textrm{H}_{2}=\textrm{MgH}
\end{equation}
Between the H$_{2}$-H boundary and the Mg-silicate cloud base, the MgH abundance is proportional to $P_{T}^{0.5}$ and $m^{1}$ (see Table \ref{table magnesium reactions}). In contrast, the MgH
abundance above the magnesium-silicate clouds is proportional to $P_{T}^{-0.5}$ and $m^{-1}$.  The sharp bends in the MgH abundance contours occur at the condensation temperature of forsterite.
Slight inflections in the MgH abundance occur along the CH$_{4}$-CO boundary, because the Mg abundance in reaction (\ref{reaction Mg:MgH}) is sensitive to the H$_{2}$O abundance in reactions
(\ref{reaction forsterite condensation}) and (\ref{reaction enstatite condensation}). Magnesium hydride is an important opacity source from 0.44 to 0.56 $\mu$m \citep{weck2003}, and MgH bands at
0.48 and 0.52 $\mu$m have been found in the optical spectra of bright L dwarfs and extreme subdwarfs \citep{kirkpatrick1999, reid2000, kirkpatrick2005ARAA}.

\subsubsection{Magnesium Monohydroxide, MgOH}\label{sss Magnesium monohydroxide, MgOH}
The chemical behavior of MgOH as a function of pressure and temperature is shown in Figure \ref{figure magnesium gas 1}c.  The equilibrium
abundance of MgOH is governed by the reaction
\begin{equation}\label{reaction Mg:MgOH}
\textrm{Mg}+\textrm{H}_{2}\textrm{O}=\textrm{MgOH}+0.5\textrm{H}_{2},
\end{equation}
decreasing with temperature and proportional to $P_{T}^{0.5}$ and $m^{2}$ below the Mg-silicate clouds.  The kinks in the MgOH contours occur where forsterite condenses.  Above the clouds,
$X_{\textrm{\scriptsize{MgOH}}}$ is proportional to $P_{T}^{-0.5}$ (see Table \ref{table magnesium reactions}).  Interestingly, the MgOH abundance in this region is independent of metallicity
and shows no shift at the CH$_{4}$-CO boundary because these effects cancel out in reaction (\ref{reaction Mg:MgOH}). For example, the H$_{2}$O abundance slightly decreases when moving from the
CH$_{4}$ to the CO field \citep[e.g.][]{lodders2002}. However, as can be seen by comparing equations (\ref{equation Mg above clouds CH4}) and (\ref{equation Mg above clouds CO}), there is a
corresponding increase in the Mg abundance, so the resulting MgOH abundance in reaction (\ref{reaction Mg:MgOH}) remains unaffected by carbon chemistry. Magnesium monohydroxide is typically the
third most abundant Mg-bearing gas in substellar atmospheres.

\subsubsection{Magnesium Hydroxide, Mg(OH)$_{2}$}\label{sss Magnesium Hydroxide, Mg(OH)2}
Mole fraction abundance contours for Mg(OH)$_{2}$ are shown in Figure \ref{figure magnesium gas 1}d for a solar-metallicity gas.  The equilibrium conversion between Mg and Mg(OH)$_{2}$ is
represented by reaction (\ref{reaction Mg:Mg(OH)2}).  Below the Mg-silicate clouds, the Mg(OH)$_{2}$ abundance decreases with temperature and pressure, and has a very strong ($m^{3}$) dependence
on metallicity.

Above the magnesium-silicate clouds (the kinks in Figure \ref{figure magnesium gas 1}d), the Mg(OH)$_{2}$ abundance contours show inflections along the CH$_{4}$-CO boundary, because Mg(OH)$_{2}$
is sensitive to the water abundance in reaction (\ref{reaction Mg:Mg(OH)2}).  Within either the CH$_{4}$ or CO field, the Mg(OH)$_{2}$ abundance is pressure-independent.   In principle, the
abundance of Mg(OH)$_{2}$ is therefore diagnostic of atmospheric temperature if the object's metallicity is known.

At low temperatures and high pressures, Mg(OH)$_{2}$ becomes the most abundant Mg-bearing gas, as illustrated in Figure \ref{figure magnesium chemistry overview}.  Magnesium hydroxide is thus
expected to be the dominant magnesium gas below $\sim$1550 K in the atmosphere of Jupiter \citep[cf.][]{fegley1994} and below $\sim$980 K on Gliese 229B.

\subsubsection{Magnesium Monoxide, MgO}\label{sss Magnesium Monoxide, MgO}
\begin{figure}
\scalebox{.7}{\includegraphics{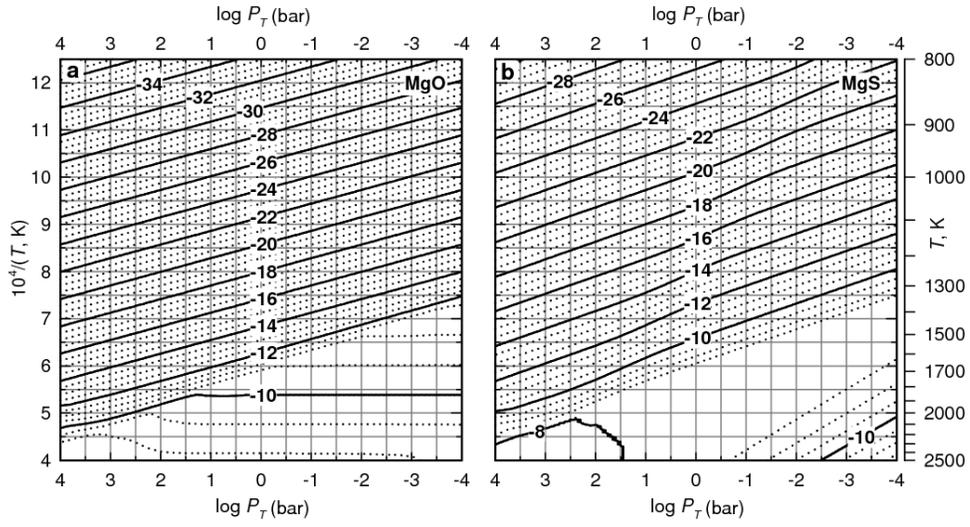}} \caption[Individual Mg-bearing gases 2]{Mole fraction contours (on a logarithmic scale) for (a) magnesium monoxide (MgO) and (b) magnesium sulfide (MgS)
as a function of pressure and temperature in a solar-metallicity gas.} \label{figure magnesium gas 2}
\end{figure}

The abundance of MgO as a function of pressure and temperature is illustrated in Figure \ref{figure magnesium gas 2}a.  Magnesium monoxide forms
via the net thermochemical reaction
\begin{equation}\label{reaction Mg:MgO}
\textrm{Mg}+\textrm{H}_{2}\textrm{O}=\textrm{MgO}+\textrm{H}_{2}
\end{equation}
Below the Mg-silicate cloud deck, at pressures less than $\sim10$ bar, the MgO abundance is effectively pressure-independent and is therefore
potentially diagnostic of atmospheric temperature if the metallicity is known.

For similar reasons as for MgOH (see \S\ref{sss Magnesium monohydroxide, MgOH}), the MgO abundance above the magnesium-silicate clouds is
independent of metallicity and is unaffected by the CH$_{4}$-CO boundary. For example, the Mg abundance decreases with metallicity, as shown by
equations (\ref{equation Mg above clouds CH4}) and (\ref{equation Mg above clouds CO}), whereas the H$_{2}$O abundance increases with
metallicity. These effects cancel out in reaction (\ref{reaction Mg:MgO}), with the result that MgO is unaffected by changes in metallicity or
the prevailing carbon chemistry.

\subsubsection{Magnesium Sulfide, MgS}\label{sss Magnesium Sulfide, MgS}
Figure \ref{figure magnesium gas 2}b displays mole fraction contours for MgS in protosolar composition gas.  The equilibrium abundance of MgS is
governed by the reaction
\begin{equation}\label{reaction Mg:MgS}
\textrm{Mg}+\textrm{H}_{2}\textrm{S}=\textrm{MgS}+\textrm{H}_{2}
\end{equation}
Near the silicate cloud base, the MgS abundance is $\sim10$ ppb in a solar-metallicity gas and decreases at higher temperatures and lower pressures as H$_{2}$S is replaced by monatomic S.  Above
the Mg-silicate clouds, $X_{\textrm{\scriptsize{MgS}}}$ is proportional to $P_{T}^{-1}$ and is metallicity-independent (see Table \ref{table magnesium reactions}). Magnesium sulfide is typically
among the more abundant Mg-bearing gases in brown dwarfs and extrasolar giant planets, and becomes the second most abundant magnesium gas at temperatures below $\sim$1600 K in the atmospheres of
L dwarfs and Pegasi planets.

\subsection{Magnesium Chemistry in Substellar Objects}\label{ss Magnesium Chemistry in Substellar Objects}

\begin{figure}
\scalebox{.7}{\includegraphics{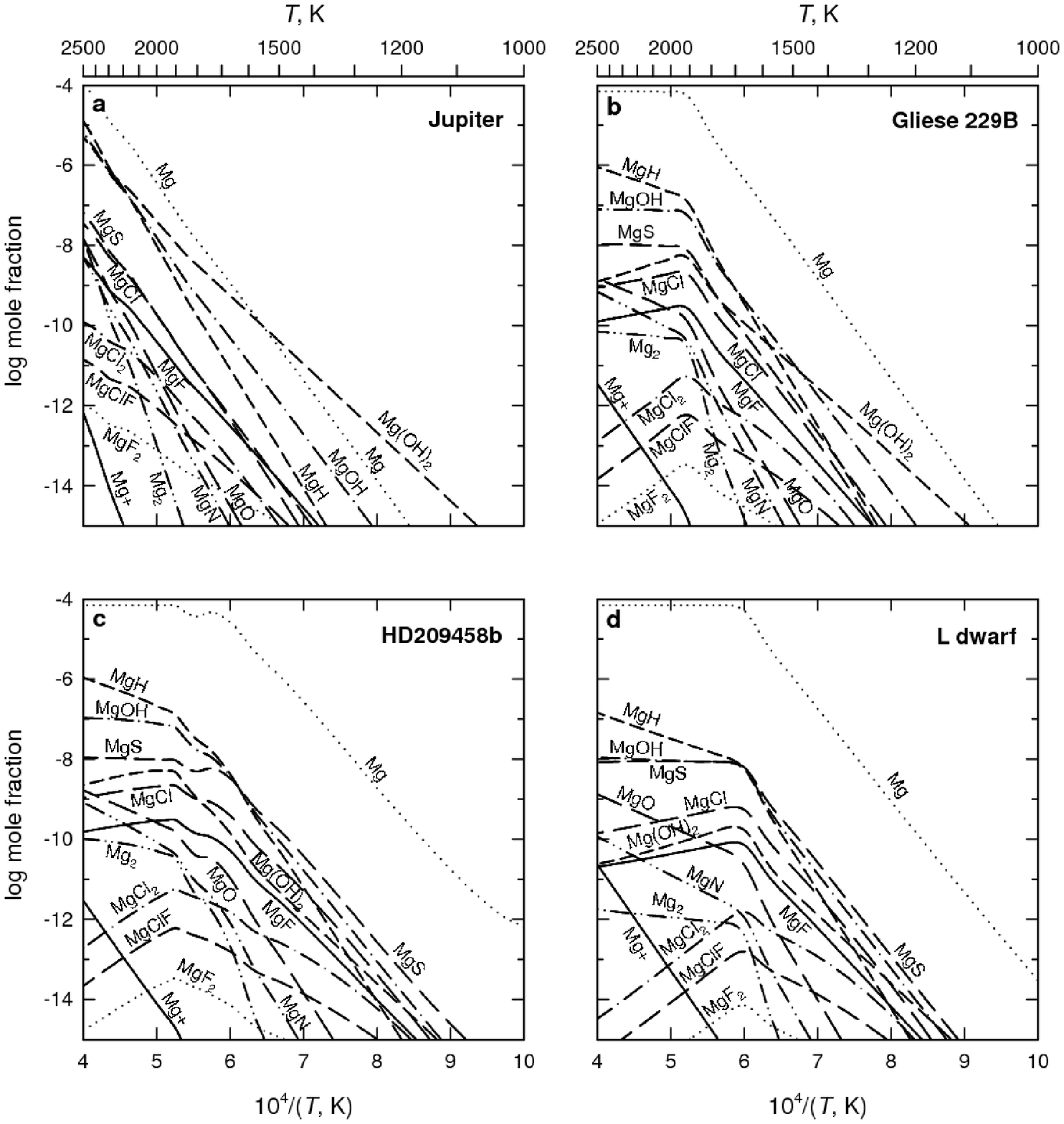}} \caption[Summary of magnesium chemistry]{Magnesium chemistry along the atmospheric pressure-temperature profile from 1000 to 2500 K for (a) Jupiter, (b)
Gliese 229B, (c) HD209458b, and (d) an L dwarf ($T_{\textrm{\scriptsize{eff}}}=1800$ K).  Breaks in the Mg curves show where magnesium-silicates condense (b,c,d).  Forsterite is condensed
throughout the entire temperature range shown here for Jupiter (a).} \label{figure magnesium chemistry summary}
\end{figure}

Magnesium gas chemistry along the pressure-temperature profiles of Jupiter, Gliese 229B, HD209458b, and an L dwarf ($T_{\textrm{\scriptsize{eff}}}=1800$ K) are illustrated in Figures \ref{figure
magnesium chemistry summary}a-\ref{figure magnesium chemistry summary}d.  Monatomic Mg is the dominant Mg-bearing gas throughout the atmospheres of Pegasi planets and L dwarfs, and in the deep
atmospheres of giant planets and T dwarfs.  Furthermore, the Mg abundance below the Mg-silicate cloud deck is essentially constant and representative of the total Mg abundance in the atmosphere.
Upon condensation, the abundances of magnesium-bearing gases rapidly decrease with decreasing temperature above the magnesium-silicate clouds. In objects with lower effective temperatures, a
number of other Mg-bearing gases become relatively abundant and Mg(OH)$_{2}$ replaces Mg as the most abundant magnesium gas at the $\sim1550$ K level on Jupiter and the $\sim980$ K level on
Gliese 229B. Magnesium hydride (MgH) is the second most abundant magnesium gas in the deep atmospheres of substellar objects until it is replaced at lower temperatures by Mg(OH)$_{2}$ and MgOH
(in giant planets and T dwarfs) or MgS (in L dwarfs and Pegasi planets).  Magnesium hydroxide (Mg(OH)$_{2}$) and MgO are potential atmospheric temperature probes, respectively, above and below
the magnesium-silicate clouds.

\section{Silicon Chemistry}\label{s Silicon Chemistry}
\subsection{Overview of Silicon Chemistry}\label{ss Overview of Silicon Chemistry}
An overview of silicon chemistry as a function of pressure and temperature is illustrated in Figure \ref{figure silicon chemistry overview}.  The dotted lines labeled Mg$_{2}$SiO$_{4}$(s,l) and
MgSiO$_{3}$(s,l) show the condensation temperatures of forsterite and enstatite, which together remove nearly all silicon from the overlying atmosphere. \citet{cushing2006} found that silicate
absorption features near 10 $\mu$m in mid-type L dwarf spectra are consistent with the presence of these magnesium silicates, but noted the possibility of additional absorption by quartz
(SiO$_{2}$) grains based upon the predictions of \citet{helling2006}.  In contrast to the models of \citet{helling2006,helling2008}, we find that quartz will not condense in the atmospheres of
substellar objects unless enstatite condensation is suppressed. This is demonstrated in Figure 10, which shows the elemental distribution of Si in condensed phases at 1 bar total pressure in a
solar-metallicity gas with (Figure \ref{figure enstatite}a) and without (Figure \ref{figure enstatite}b) enstatite condensation.  As shown in Figure \ref{figure enstatite},  Mg$_{2}$SiO$_{4}$
formation consumes nearly half of the total Si abundance because the solar elemental abundances of Mg and Si are approximately equal. Enstatite formation plausibly proceeds via reactions between
SiO gas and pre-existing forsterite grains and continues until nearly all silicon is consumed. Thus, quartz ($T_{\textrm{\scriptsize{cond}}}\sim 1550$ K) can only form in the absence of
enstatite ($T_{\textrm{\scriptsize{cond}}}\sim 1600$ K), because MgSiO$_{3}$ otherwise efficiently removes silicon from the gas phase.  Even in the absence of gas-grain reactions between SiO and
Mg$_{2}$SiO$_{4}$, the vapor pressures of Mg and SiO above forsterite (reaction \ref{reaction forsterite condensation}) remain high enough to drive enstatite cloud formation via the net
thermochemical reaction (\ref{reaction enstatite condensation}), so that MgSiO$_{3}$ condenses instead of SiO$_{2}$.  We therefore conclude that SiO$_{2}$ will not condense within the silicate
cloud.

The most abundant Si-bearing gas over a wide range of pressures and temperatures is SiO, until it is replaced at higher pressures by silane, SiH$_{4}$ (see Figure \ref{figure silicon chemistry
overview}).  The equilibrium conversion between SiO and SiH$_{4}$ is
\begin{equation}\label{reaction SiH4:SiO}
\textrm{SiH}_{4}+\textrm{H}_{2}\textrm{O}=\textrm{SiO}+3\textrm{H}_{2}.
\end{equation}
The position of the line where SiO and SiH$_{4}$ have equal abundances A(SiO) = A(SiH$_{4}$) $\approx \frac{1}{2}\Sigma
\textrm{Si}_{\textrm{\scriptsize{gas}}}$ (where $\Sigma \textrm{Si}_{\textrm{\scriptsize{gas}}}$ is the total amount of silicon in the gas) is
given by
\begin{equation}\label{boundary line SiH4:SiO}
\log P_{T}=4.68-3086/T+0.5[\textrm{Fe/H}].
\end{equation}
As temperatures decrease, SiH$_{4}$ is replaced by SiH$_{3}$F via the net thermochemical reaction
\begin{equation}\label{reaction SiH4:SiH3F}
\textrm{SiH}_{4}+\textrm{HF}=\textrm{SiH}_{3}\textrm{F}+\textrm{H}_{2}.
\end{equation}
This reaction is independent of pressure and the position of the SiH$_{4}$-SiH$_{3}$F boundary is given by
\begin{equation}\label{boundary line SiH4:SiH3F}
T=6948/(7.24-[\textrm{Fe/H}]),
\end{equation}
and occurs at $T \sim$ 960 K in a solar-metallicity gas. Meanwhile, The equilibrium between SiO and SiH$_{3}$F is
\begin{equation}\label{reaction SiO:SiH3F}
\textrm{SiO}+\textrm{HF}+2\textrm{H}_{2}=\textrm{SiH}_{3}\textrm{F}+\textrm{H}_{2}\textrm{O},
\end{equation}
where the position of the SiO-SiH$_{3}$F line is given by
\begin{equation}\label{boundary line SiO:SiH3F}
\log P_{T} = 8.30 - 6560/T,
\end{equation}
independent of the metallicity.  The SiH$_{4}$-SiO, SiH$_{4}$-SiH$_{3}$F, and SiO-SiH$_{3}$F equal abundance lines intersect to form a ``triple point'' at $T \sim$ 960 K and $P_{T} \sim
10^{1.47}$ bar in solar-metallicity gas, where all three gases have equal abundances [A(SiO) = A(SiH$_{4}$) = A(SiH$_{3}$F) $\approx\frac{1}{3}\Sigma\textrm{Si}_{\textrm{\scriptsize{gas}}}$].

At lower temperatures, SiH$_{2}$F$_{2}$ replaces SiH$_{3}$F via the net thermochemical reaction
\begin{equation}\label{reaction SiH3F:SiH2F2}
\textrm{SiH}_{3}\textrm{F}+\textrm{HF}=\textrm{SiH}_{2}\textrm{F}_{2}+\textrm{H}_{2}.
\end{equation}
This reaction is also independent of pressure and the SiH$_{3}$F-SiH$_{2}$F$_{2}$ boundary is given by
\begin{equation}\label{boundary line SiH3F:SiH2F2}
T=7040/(7.68-[\textrm{Fe/H}]),
\end{equation}
and is located at $T \sim$ 917 K in a solar-metallicity gas.  The conversion between SiO and SiH$_{2}$F$_{2}$ takes place by the reaction
\begin{equation}\label{reaction SiO:SiH2F2}
\textrm{SiO}+\textrm{HF}+\textrm{H}_{2}=\textrm{SiH}_{2}\textrm{F}_{2}+\textrm{H}_{2}\textrm{O},
\end{equation}
where the position of the SiO-SiH$_{2}$F$_{2}$ boundary is given by
\begin{equation}\label{boundary line SiO:SiH2F2}
\log P_{T} = 12.14 - 10080/T - 0.5[\textrm{Fe/H}].
\end{equation}
In a solar-metallicity gas, equations (\ref{boundary line SiO:SiH3F}) and (\ref{boundary line SiO:SiH2F2}) intersect to form the SiO-SiH$_{3}$F-SiH$_{2}$F$_{2}$ ``triple point'' at $T \sim 917$
K and $P_{T} \sim 10^{1.15}$ bar, where all three gases have equal abundances ($X\approx10^{-19.4}$ for [Fe/H] = 0).  However, the abundances of SiH$_{3}$F and SiH$_{2}$F$_{2}$ in this region
are extremely low because most silicon is removed from the atmosphere by cloud formation at deeper levels.

\subsection{Chemical Behavior of Silicon-bearing Gases}\label{ss Chemical Behavior of Silicon-bearing Gases}
\subsubsection{Silicon Monoxide, SiO}\label{sss Silicon Monoxide, SiO}

The chemical behavior of SiO in a protosolar composition gas is illustrated in Figure \ref{figure SiO} as a function of pressure and temperature.  Within the SiO field, the SiO abundance below
the magnesium-silicate clouds is given by
\begin{equation}\label{equation SiO below clouds}
\log X_{\textrm{\scriptsize{SiO}}}\approx -4.20 + [\textrm{Fe/H}],
\end{equation}
and SiO contains $\sim$100\% of the atmospheric silicon inventory ($X_{\textrm{\scriptsize{SiO}}}\approx X_{\Sigma\textrm{\scriptsize{Si}}}$).
 Upon Mg-silicate condensation, silicon is efficiently removed from the gas phase and the SiO abundance rapidly decreases with decreasing
temperature. Above the clouds, the SiO abundance is governed by its vapor pressure over rock. Curvature in the SiO contour lines along the
CH$_{4}$-CO equal abundance boundary results from the effect of carbon chemistry on the H$_{2}$O abundance in reactions (\ref{reaction
forsterite condensation}) and (\ref{reaction enstatite condensation}). The SiO abundance is thus
\begin{equation}\label{equation SiO above clouds CH4}
X_{\textrm{\scriptsize{SiO}}}\approx 13.21 - 28817/T - \log P_{T}-[\textrm{Fe/H}],
\end{equation}
in CH$_{4}$-dominated objects and
\begin{equation}\label{equation SiO above clouds CO}
X_{\textrm{\scriptsize{SiO}}}\approx 13.67 - 28817/T - \log P_{T}-[\textrm{Fe/H}],
\end{equation}
in CO-dominated objects, inversely proportional to pressure and metallicity.  These expressions, along with chemical equilibria, are used to determine the equilibrium abundances of other
Si-bearing gases.  Expressions giving the abundances of silicon species as a function of temperature, pressure, and metallicity in a protosolar composition gas are listed in Table \ref{table
silicon reactions}.  We note that the types of cloud condensates present will affect the gas chemistry of SiO and subsequently all other Si-bearing gases.  For example, as shown in Figure
\ref{figure SiO condense}, SiO mole fraction abundances above the clouds are $\sim$0.5 dex higher if enstatite formation is suppressed and replaced by SiO$_{2}$ condensation (see \S\ref{ss
Overview of Silicon Chemistry} above).  The SiO abundance is therefore potentially diagnostic of weather and cloud composition in brown dwarf atmospheres.

Silicon monoxide has not yet been detected in the atmospheres of brown dwarfs or giant planets, but has been observed in numerous objects including molecular clouds, circumstellar envelopes, the
photospheres of late-type stars, and sunspots \citep[e.g.,][and references therein]{campbell1995}.  Abundant SiO gas was recently detected in the circumstellar disk of the $\beta$ Pic analog
HD172555 \citep{lisse2009}.

\subsubsection{Silicon Monosulfide, SiS}\label{sss Silicon Monosulfide, SiS}

Mole fraction contours for SiS are shown in Figure \ref{figure silicon gas 1}a for a solar-metallicity gas.  Silicon monosulfide is formed by the reaction between SiO and H$_{2}$S:
\begin{equation}\label{reaction SiO:SiS}
\textrm{SiO}+\textrm{H}_{2}\textrm{S}=\textrm{SiS}+\textrm{H}_{2}\textrm{O}
\end{equation}
Below the Mg-silicate cloud deck, $X_{\textrm{\scriptsize{SiS}}}\sim5$-10 ppm in a solar-metallicity gas.  The SiS abundance decreases at higher temperatures and lower pressures as H$_{2}$S is
replaced by SH and monatomic S \citep[e.g.][]{visscher2006}.

Above the clouds, the SiS abundance is inversely proportional to metallicity and total pressure (see Table \ref{table silicon reactions}). The SiS abundance contours in Figure \ref{figure
silicon gas 1}a also display curvature along the CH$_{4}$-CO boundary.  This shift is more pronounced for SiS than for SiO because the SiS abundance in reaction (\ref{reaction SiO:SiS}) depends
on the SiO and H$_{2}$O abundances, both of which are affected by carbon chemistry. For example, when reaction (\ref{reaction SiO:SiS}) is at equilibrium, LeCh\^{a}telier's principle shows that
more SiS is produced either by adding SiO (or H$_{2}$S) or removing H$_{2}$O.  When moving from the CH$_{4}$ field to the CO field in a protosolar composition gas, the SiO abundance increases
and the H$_{2}$O abundance decreases, yielding a correspondingly large increase in the SiS abundance. Silicon sulfide is expected to be the second most abundant silicon-bearing gas (after SiO)
in the atmospheres of brown dwarfs and extrasolar giant planets. Furthermore, it is a potential tracer of weather in these objects because its maximum abundance is typically achieved near the
magnesium-silicate cloud base.

\subsubsection{Monatomic Silicon, Si}\label{sss Monatomic Silicon, Si}

Figure \ref{figure silicon gas 1}b illustrates the chemical behavior of monatomic Si gas as a function of pressure and temperature.  The abundance of Si is governed by the net thermochemical
reaction
\begin{equation}\label{reaction SiO:Si}
\textrm{SiO}+\textrm{H}_{2}=\textrm{Si}+\textrm{H}_{2}\textrm{O}.
\end{equation}
Below the Mg-silicate cloud deck and at pressures less than $\sim10$ bar, the Si abundance is effectively independent of pressure and
metallicity and thus potentially diagnostic of atmospheric temperature.  Above the clouds, the Si abundance rapidly decreases with decreasing
temperature and is proportional to $P_{T}^{-1}$ and $m^{-2}$ (see Table \ref{table silicon reactions}).  A number of Si absorption bands are
observed in the near infrared spectra of low-mass dwarf stars. These features generally weaken and disappear in late-type M dwarfs
\citep{cushing2005}, consistent with the removal of silicon into Mg$_{2}$SiO$_{4}$ and MgSiO$_{3}$ clouds deeper in the atmosphere.

\subsubsection{Silylidyne, SiH}\label{sss Silylidyne, SiH} The abundance of SiH as a function of pressure and temperature is
illustrated in Figure \ref{figure silicon gas 2}a.  The SiH abundance is governed by the net thermochemical reaction
\begin{equation}\label{reaction SiO:SiH}
\textrm{SiO}+1.5\textrm{H}_{2}=\textrm{SiH}+\textrm{H}_{2}\textrm{O}
\end{equation}
Below the Mg-silicate clouds, the SiH abundance is proportional to $P_{T}^{0.5}$ and is independent of metallicity.  Above the clouds, the SiH abundance is proportional to $P_{T}^{-0.5}$ and
$m^{-2}$ (see Table \ref{table silicon reactions}).  When moving from the CH$_{4}$ to the CO field, the SiO abundance increases and the H$_{2}$O abundance decreases in reaction (\ref{reaction
SiO:SiH}).  Both effects serve to increase the SiH abundance, resulting in a shift in the SiH contour lines along the CH$_{4}$-CO boundary.  The $m^{-2}$ dependence on metallicity means that for
every $[\textrm{Fe/H}]=+1$ dex increase in metallicity, the SiH abundance decreases by a factor of 100 at a given pressure and temperature.

\subsubsection{Silylene, SiH$_{2}$}\label{sss Silylene, SiH2}
Mole fraction contours for SiH$_{2}$ are shown in Figure \ref{figure silicon gas 2}b.  The equilibrium between SiO and SiH$_{2}$ is given by the
reaction
\begin{equation}\label{reaction SiO:SiH2}
\textrm{SiO}+2\textrm{H}_{2}=\textrm{SiH}_{2}+\textrm{H}_{2}\textrm{O}
\end{equation}
Beneath the magnesium-silicate cloud deck, the metallicity-independent silylene abundance increases with total pressure. Upon rock condensation,
the amount of SiH$_{2}$ remaining in the gas rapidly decreases with decreasing temperature. As for SiH, the SiH$_{2}$ abundance is sensitive to
the SiO and H$_{2}$O abundances in reaction and thus a shift occurs in the SiH$_{2}$ contour lines when moving between the CH$_{4}$ and CO
fields. Within each field, the SiH$_{2}$ abundance is pressure-independent (see Figure \ref{figure silicon gas 2} and Table \ref{table silicon
reactions}), and thus potentially diagnostic of temperature if the metallicity is known and SiH$_{2}$ is thermochemical in origin.

\subsubsection{Silyl Radical, SiH$_{3}$}\label{sss Silyl Radical, SiH3}
The chemical behavior of the SiH$_{3}$ radical in a solar-metallicity gas is illustrated in Figure \ref{figure silicon gas 2}c.  The abundance of SiH$_{3}$ is governed by the reaction
\begin{equation}\label{reaction SiO:SiH3}
\textrm{SiO}+2.5\textrm{H}_{2}=\textrm{SiH}_{3}+\textrm{H}_{2}\textrm{O}
\end{equation}
Between the H$_{2}$-H boundary and the Mg-silicate cloud base, the SiH$_{3}$ is proportional to $P_{T}^{1.5}$.  Above the clouds, the SiH$_{3}$ abundance is proportional to $P_{T}^{1.5}$ and
$m^{-2}$ (see Table \ref{table silicon reactions}). There is curvature in the SiH$_{3}$ contour lines along the CH$_{4}$-CO boundary because the prevailing carbon chemistry affects both the SiO
and H$_{2}$O abundances in reaction (\ref{reaction SiO:SiH3})

\subsubsection{Silane, SiH$_{4}$}\label{sss Silane, SiH4}
Mole fraction abundance contours for SiH$_{4}$ are illustrated in Figure \ref{figure silicon gas 2}d. The equilibrium conversion between SiO and SiH$_{4}$ is represented by reaction
(\ref{reaction SiH4:SiO}), and expressions giving the SiH$_{4}$ abundance as a function of pressure, temperature, and metallicity are listed in Table \ref{table silicon reactions}.  Below the
clouds, the silane abundance has a strong ($P_{T}^{2}$) dependence on total pressure. Upon Mg-silicate cloud formation, there is an inflection in the SiH$_{4}$ contour lines, and the SiH$_{4}$
abundance above the clouds is proportional to $P_{T}$ and $m^{-2}$.  Curvature in the SiH$_{4}$ contour lines along the CH$_{4}$-CO boundary results from the effect of carbon chemistry on the
SiO and H$_{2}$O abundances in reaction (\ref{reaction SiH4:SiO}).

At high pressures and temperatures ($>960$ K for $[\textrm{Fe/H}]=0$), SiH$_{4}$ becomes the most abundant silicon-bearing gas.  We thus expect SiH$_{4}$ to be the dominant silicon gas below the
1031 K level in the deep atmosphere of Jupiter ($[\textrm{Fe/H}]\approx+0.5$).  However, we again emphasize the efficiency with which rock condensation removes silicon from the gas, and we
expect an abundance of $X_{\textrm{\scriptsize{SiH}}_{4}}\approx 10^{-16.5}$ at the 1031 K level on Jupiter.  At lower temperatures, SiH$_{4}$ is surpassed by SiH$_{3}$F and SiH$_{2}$F$_{2}$.
Silane is expected to be the third most abundant Si-bearing gas in the deep atmosphere of Gliese 229B.

We also point out the trend in pressure dependence for the silicon hydrides below ($X_{\textrm{\scriptsize{SiH}}}\propto P_{T}^{0.5}$, $X_{\textrm{\scriptsize{SiH}}_{2}}\propto P_{T}^{1}$,
$X_{\textrm{\scriptsize{SiH}}_{3}}\propto P_{T}^{1.5}$, $X_{\textrm{\scriptsize{SiH}}_{4}}\propto P_{T}^{2}$) and above the silicate clouds ($X_{\textrm{\scriptsize{SiH}}}\propto P_{T}^{-0.5}$,
$X_{\textrm{\scriptsize{SiH}}_{2}}\propto P_{T}^{0}$, $X_{\textrm{\scriptsize{SiH}}_{3}}\propto P_{T}^{0.5}$, $X_{\textrm{\scriptsize{SiH}}_{4}}\propto P_{T}^{1}$), which is evident in the
shapes of the contour lines in Figure \ref{figure silicon gas 2}. The abundance of each of the silicon hydrides is metallicity-independent below the Mg-silicate clouds and proportional to
$m^{-2}$ above the clouds.

\subsection{Silicon Chemistry in Substellar Atmospheres}\label{ss Silicon Chemistry in Substellar Atmospheres}

Figures \ref{figure silicon chemistry summary}a-\ref{figure silicon chemistry summary}d summarize the chemical behavior of silicon gases along the pressure temperature profiles of Jupiter, the T
dwarf Gliese 229B, the Pegasi planet HD209458b, and an L dwarf ($T_{\textrm{\scriptsize{eff}}}=1800$ K).  There is a clear trend in silicon chemistry as a function of effective temperature.  On
Jupiter, SiH$_{4}$ is the dominant Si-bearing gas throughout the deep atmosphere, and is replaced by SiH$_{3}$F and SiH$_{2}$F$_{2}$ at lower temperatures (see Figure \ref{figure silicon
chemistry overview}).  The second most abundant Si-bearing gas is SiO, followed by SiS and a number of other silicon gases.  In the warmer atmosphere of Gliese 229B, SiO is the dominant
Si-bearing gas, followed by SiS and SiH$_{4}$.  The relative importance of SiH$_{4}$ decreases with increasing effective temperature, and SiO and SiS are the most important silicon gases
throughout the atmospheres of Pegasi planets and L dwarfs.  The abundances of all the silicon gases shown in Figure \ref{figure silicon chemistry summary} rapidly decrease with decreasing
temperature above the silicate clouds, which explains the non-detection of SiH$_{4}$ and other silicon species in the atmospheres of Jupiter and Saturn. Below the cloud base, the important
silicon gases SiO and SiS reach their maximum abundance and are therefore potential tracers of weather in brown dwarfs and Pegasi planets.   Silylene (SiH$_{2}$) and monatomic silicon (Si) are
potentially diagnostic of atmospheric temperature, respectively, above and below the magnesium silicate clouds.  The abundant silicon gases SiO and SiS typically achieve their maximum abundance
near the cloud base and are potential tracers of weather in brown dwarfs and Pegasi planets.

\section{Summary}\label{s Summary}

The chemical behavior of iron species in substellar atmospheres is strongly affected by Fe metal condensation, which efficiently removes most iron from the gas phase.  Similarly, most magnesium
and silicon is removed from the gas by forsterite (Mg$_{2}$SiO$_{4}$) and enstatite (MgSiO$_{3}$) cloud formation.  The equilibrium abundances of Fe-, Mg-, and Si-bearing gases rapidly decrease
with increasing altitude (and decreasing temperature) above the clouds.

Monatomic iron is the dominant Fe-bearing gas throughout the atmospheres of L dwarfs and Pegasi planets.  Other less abundant iron gases become increasingly important in objects with lower
effective temperatures, and Fe(OH)$_{2}$ replaces Fe at low temperatures in T dwarfs and giant planets. Magnesium gas chemistry is similar to that of iron. Monatomic Mg is the most abundant
magnesium gas throughout the atmospheres of L dwarfs and Pegasi planets and in the deep atmospheres of giant planets and T dwarfs, where Mg is replaced by Mg(OH)$_{2}$ at lower temperatures. A
number of Mg-bearing gases become relatively abundant with decreasing effective temperature. Silicon monoxide (SiO) is the most abundant Si-bearing gas, followed by SiS, throughout the
atmospheres of L dwarfs and Pegasi planets and in the deep atmospheres of T dwarfs.  In objects with lower effective temperatures, a number of other silicon gases become increasingly important
and SiH$_{4}$ is the dominant silicon gas in the deep atmosphere of Jupiter.  At high pressures and low temperatures SiH$_{4}$ and SiO are replaced by SiH$_{3}$F and/or SiH$_{2}$F$_{2}$.

The abundances of several Fe-, Mg-, and Si-bearing gases are pressure-independent and thus, in principle, diagnostic of atmospheric temperature. These include Fe(OH)$_{2}$, Mg(OH)$_{2}$, and Si
below the clouds and FeO, MgO, and SiH$_{2}$ above the clouds.  In addition, a number of gases (e.g. Fe, FeH, FeOH, FeS, Mg, MgH, MgOH, MgS, SiO, SiS) may serve as indicators of weather since
they generally reach their maximum abundance just below the iron metal or magnesium-silicate cloud decks.  This may be particularly useful for late M dwarfs and early L dwarfs in which the metal
and silicate clouds are located at relatively shallow depths below the photosphere.

\acknowledgments This research was conducted at Washington University in St.~Louis and was supported by the NASA Planetary Atmospheres Program (NNG06GC26G). Support for K.~Lodders was also
provided by NSF Grant AST-0707377.  Final preparation of the manuscript was supported by the Lunar and Planetary Institute/USRA (NASA Cooperative Agreement NCC5-679). LPI Contribution No. XXXX.

\begin{deluxetable}{llc}
\tablewidth{0pt}\tablecolumns{3}\tablecaption{Iron Gas Abundances} \tablehead{\colhead{gas M} & \colhead{log
$X_{\textrm{\scriptsize{M}}}\approx$} & \colhead{reaction no.}}\startdata \sidehead{Below iron cloud}
Fe & $-4.24 + [\textrm{Fe/H}]$ & \nodata\\
FeH & $-6.13 - 1905/T + 0.5 \log P_{T} + [\textrm{Fe/H}]$ & \ref{reaction Fe:FeH}\\
FeOH & $-9.80 + 2468/T + 0.5 \log P_{T} + 2[\textrm{Fe/H}]$ & \ref{reaction Fe:FeOH}\\
Fe(OH)$_{2}$ & $-16.64 + 12088/T + \log P_{T} + 3[\textrm{Fe/H}]$ & \ref{boundary line Fe:Fe(OH)2}\\
FeO & $-7.20 - 4713/T + 2[\textrm{Fe/H}]$ & \ref{reaction Fe:FeO}\\
FeS & $-8.52 + 964/T + 2[\textrm{Fe/H}]$ & \ref{reaction Fe:FeS}\\
\sidehead{Above iron cloud}
Fe & $7.23 - 20995/T - \log P_{T}$ & \ref{reaction Fe condensation}\\
FeH & $5.34 - 22900/T - 0.5 \log P_{T}$ & \ref{reaction Fe:FeH}\\
FeOH & $5.02 - 18527/T - 0.5 \log P_{T} + 2[\textrm{Fe/H}] + \log X_{\textrm{\scriptsize{H}}_{2}\textrm{\scriptsize{O}}}'$ & \ref{reaction Fe:FeOH}\\
Fe(OH)$_{2}$ & $1.53 - 8907/T + 2[\textrm{Fe/H}] + 2\log X_{\textrm{\scriptsize{H}}_{2}\textrm{\scriptsize{O}}}'$ & \ref{reaction Fe:Fe(OH)2}\\
FeO & $7.62 - 25708/T -\log P_{T} + [\textrm{Fe/H}] + \log X_{\textrm{\scriptsize{H}}_{2}\textrm{\scriptsize{O}}}'$ & \ref{reaction Fe:FeO}\\
FeS & $2.96 - 20031/T -\log P_{T} + [\textrm{Fe/H}]$ & \ref{reaction Fe:FeS}\\
\enddata
\tablecomments{$X_{\textrm{\scriptsize{H}}_{2}\textrm{\scriptsize{O}}}'$ is defined as the H$_{2}$O mole fraction in a solar-metallicity gas, where $\log
X_{\textrm{\scriptsize{H}}_{2}\textrm{\scriptsize{O}}}'\approx -3.12$ within the CH$_{4}$ field and $\approx -3.58$ within the CO field. The effect of metallicity on the atmospheric water
abundance ($X_{\textrm{\scriptsize{H}}_{2}\textrm{\scriptsize{O}}}$) is included in the abundance equations.  Expressions are valid for temperatures between 800 and 2500 K and metallicities up
to [Fe/H] = +0.5 dex.} \label{table iron reactions}
\end{deluxetable}

\begin{deluxetable}{llc}
\tablewidth{0pt}\tablecolumns{3}\tablecaption{Magnesium Gas Abundances} \tablehead{\colhead{gas M} & \colhead{log
$X_{\textrm{\scriptsize{M}}}\approx$} & \colhead{reaction no.}}\startdata \sidehead{Below magnesium-silicate clouds}
Mg & $-4.15 + [\textrm{Fe/H}]$ & \nodata\\
MgH & $-5.46 - 4236/T + 0.5\log P_{T} + [\textrm{Fe/H}]$ & \ref{reaction Mg:MgH}\\
MgOH & $-8.98 + 1672/T + 0.5\log P_{T} + 2[\textrm{Fe/H}]$ & \ref{reaction Mg:MgOH}\\
Mg(OH)$_{2}$ & $-15.54 + 10267/T + \log P_{T} + 3[\textrm{Fe/H}]$ & \ref{reaction Mg:Mg(OH)2}\\
MgO & $-6.12 - 7306/T + 2[\textrm{Fe/H}]$ & \ref{reaction Mg:MgO}\\
MgS & $-7.94 + 2[\textrm{Fe/H}]$ & \ref{reaction Mg:MgS}\\
\sidehead{Above magnesium-silicate clouds}
Mg & $8.25 - 27250/T - \log P_{T} - [\textrm{Fe/H}]- \log X_{\textrm{\scriptsize{H}}_{2}\textrm{\scriptsize{O}}}'$ & \ref{reaction forsterite condensation}, \ref{reaction enstatite condensation}\\
MgH & $6.94 - 31486/T - 0.5\log P_{T} - [\textrm{Fe/H}]- \log X_{\textrm{\scriptsize{H}}_{2}\textrm{\scriptsize{O}}}'$ & \ref{reaction Mg:MgH}\\
MgOH & $6.75 - 25578/T - 0.5\log P_{T}$ & \ref{reaction Mg:MgOH}\\
Mg(OH)$_{2}$ & $3.53 - 16983/T + [\textrm{Fe/H}]+ \log X_{\textrm{\scriptsize{H}}_{2}\textrm{\scriptsize{O}}}'$ & \ref{reaction Mg:Mg(OH)2}\\
MgO & $9.63 - 34556/T - \log P_{T}$ & \ref{reaction Mg:MgO}\\
MgS & $4.40 - 27250/T - \log P_{T}- \log X_{\textrm{\scriptsize{H}}_{2}\textrm{\scriptsize{O}}}'$ & \ref{reaction Mg:MgS}\\
\enddata
\tablecomments{$X_{\textrm{\scriptsize{H}}_{2}\textrm{\scriptsize{O}}}'$ is defined as the H$_{2}$O mole fraction in a solar-metallicity gas, where $\log
X_{\textrm{\scriptsize{H}}_{2}\textrm{\scriptsize{O}}}'\approx -3.12$ within the CH$_{4}$ field and $\approx -3.58$ within the CO field.  The effect of metallicity on the atmospheric water
abundance ($X_{\textrm{\scriptsize{H}}_{2}\textrm{\scriptsize{O}}}$) is included in the abundance equations. Expressions are valid for temperatures between 800 and 2500 K and metallicities up to
[Fe/H] = +0.5 dex.} \label{table magnesium reactions}
\end{deluxetable}

\begin{deluxetable}{llc}
\tablewidth{0pt}\tablecolumns{3}\tablecaption{Silicon Gas Abundances} \tablehead{\colhead{gas M} & \colhead{log
$X_{\textrm{\scriptsize{M}}}\approx$} & \colhead{reaction no.}}\startdata \sidehead{Below magnesium-silicate clouds; within SiO field}
SiO & $-4.20 + [\textrm{Fe/H}]$ & \nodata\\
SiS & $-5.59 + 666/T + 2[\textrm{Fe/H}]$ & \ref{reaction SiO:SiS}\\
Si & $-0.44-44738/T$ & \ref{reaction SiO:Si}\\
SiH & $-2.64 - 11500/T + 0.5 \log P_{T}$ & \ref{reaction SiO:SiH}\\
SiH$_{2}$ & $-5.83 - 6422/T + \log P_{T}$ & \ref{reaction SiO:SiH2}\\
SiH$_{3}$ & $-8.97 - 2770/T + 1.5\log P_{T}$ & \ref{reaction SiO:SiH3}\\
SiH$_{4}$ & $-13.33 + 6172/T + 2\log P_{T}$ & \ref{reaction SiH4:SiO}\\
SiH$_{3}$F & $-20.57 + 13120/T + 2\log P_{T} + [\textrm{Fe/H}]$ & \ref{reaction SiO:SiH3F}\\
SiH$_{2}$F$_{2}$ & $-28.25 + 20160/T + 2\log P_{T} + 2[\textrm{Fe/H}]$ & \ref{reaction SiO:SiH2F2}\\

\sidehead{Above magnesium-silicate clouds}
SiO & $10.09 - 28817/T - \log P_{T} - [\textrm{Fe/H}] - \log X_{\textrm{\scriptsize{H}}_{2}\textrm{\scriptsize{O}}}'$ & \ref{reaction forsterite condensation}, \ref{reaction enstatite condensation}\\
SiS & $5.38 - 28151/T - \log P_{T} - [\textrm{Fe/H}] - 2\log X_{\textrm{\scriptsize{H}}_{2}\textrm{\scriptsize{O}}}'$ & \ref{reaction SiO:SiS}\\
Si & $10.45-44659/T - \log P_{T} - 2[\textrm{Fe/H}] - 2\log X_{\textrm{\scriptsize{H}}_{2}\textrm{\scriptsize{O}}}'$ & \ref{reaction SiO:Si}\\
SiH & $8.30 - 40317/T - 0.5\log P_{T} - 2[\textrm{Fe/H}] - 2\log X_{\textrm{\scriptsize{H}}_{2}\textrm{\scriptsize{O}}}'$ & \ref{reaction SiO:SiH}\\
SiH$_{2}$ & $5.11 - 35239/T - 2[\textrm{Fe/H}] - 2\log X_{\textrm{\scriptsize{H}}_{2}\textrm{\scriptsize{O}}}'$ & \ref{reaction SiO:SiH2}\\
SiH$_{3}$ & $1.97 - 31587/T + 0.5\log P_{T} - 2[\textrm{Fe/H}]- 2\log X_{\textrm{\scriptsize{H}}_{2}\textrm{\scriptsize{O}}}'$ & \ref{reaction SiO:SiH3}\\
SiH$_{4}$ & $-2.39 - 22645/T + \log P_{T} - 2[\textrm{Fe/H}]- 2\log X_{\textrm{\scriptsize{H}}_{2}\textrm{\scriptsize{O}}}'$ & \ref{reaction SiH4:SiO}\\
SiH$_{3}$F & $-9.63 - 15697/T + \log P_{T} - [\textrm{Fe/H}] - 2\log X_{\textrm{\scriptsize{H}}_{2}\textrm{\scriptsize{O}}}'$ & \ref{reaction SiO:SiH3F}\\
SiH$_{2}$F$_{2}$ & $-17.31 - 8657/T + \log P_{T}- 2\log X_{\textrm{\scriptsize{H}}_{2}\textrm{\scriptsize{O}}}'$ & \ref{reaction SiO:SiH2F2}\\
\enddata
\tablecomments{$X_{\textrm{\scriptsize{H}}_{2}\textrm{\scriptsize{O}}}'$ is defined as the H$_{2}$O mole fraction in a solar-metallicity gas, where $\log
X_{\textrm{\scriptsize{H}}_{2}\textrm{\scriptsize{O}}}'\approx -3.12$ within the CH$_{4}$ field and $\approx -3.58$ within the CO field.  The effect of metallicity on the atmospheric water
abundance ($X_{\textrm{\scriptsize{H}}_{2}\textrm{\scriptsize{O}}}$) is included in the abundance equations.  Expressions are valid for temperatures between 800 and 2500 K and metallicities up
to [Fe/H] = +0.5 dex.} \label{table silicon reactions}
\end{deluxetable}

\begin{figure}
\scalebox{.7}{\includegraphics{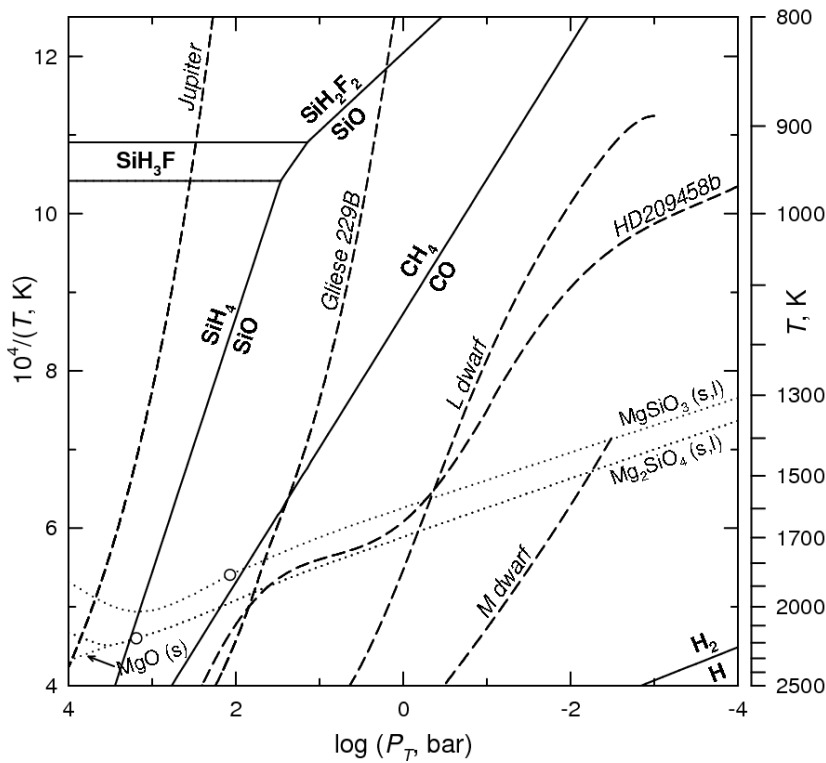}} \caption[Overview of silicon chemistry]{Overview of silicon chemistry as a function of temperature and pressure in a protosolar composition gas.  The
solid line indicates where dominant Si-bearing gases have equal abundances. The dotted lines labeled Mg$_{2}$SiO$_{4}$ and MgSiO$_{3}$ show the condensation temperatures of forsterite and
enstatite, with circles denoting their melting points. Also shown are model atmospheric profiles for representative substellar objects (dashed lines). See text for details.} \label{figure
silicon chemistry overview}
\end{figure}

\begin{figure}
\scalebox{.7}{\includegraphics{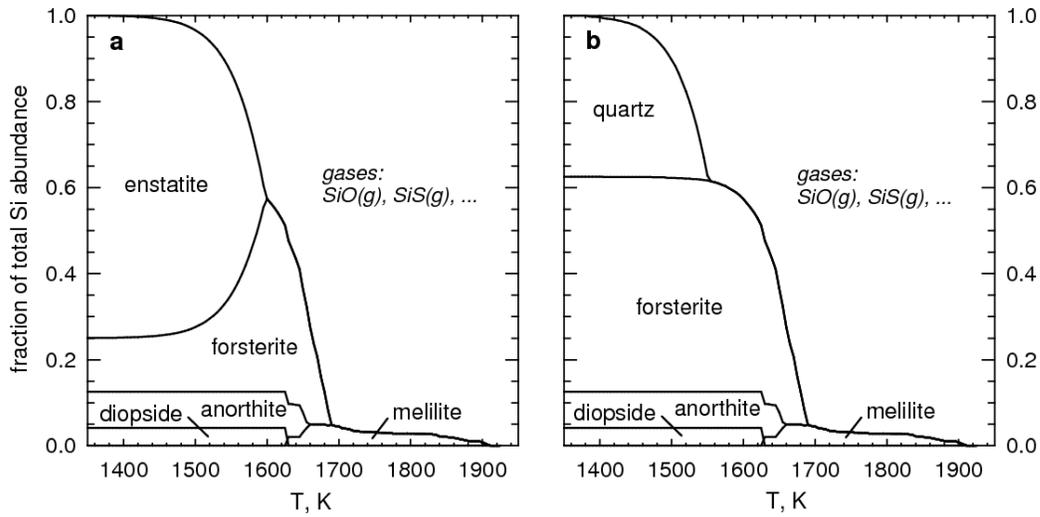}} \caption[Enstatite Condensation Chemistry]{Elemental distribution of silicon into condensed phases at 1 bar in a solar-metallicity gas with enstatite
(MgSiO$_{3}$) formation a) included or b) suppressed.  Quartz (SiO$_{2}$) condensation will only proceed in the absence of enstatite, which otherwise efficiently removes silicon from the gas
phase.  The silicates melilite, anorthite, and diopside together consume up to $\sim$12\% of the total Si abundance.  SiO and SiS are the dominant Si-bearing gases before removal by
condensation.} \label{figure enstatite}
\end{figure}

\begin{figure}
\scalebox{.7}{\includegraphics{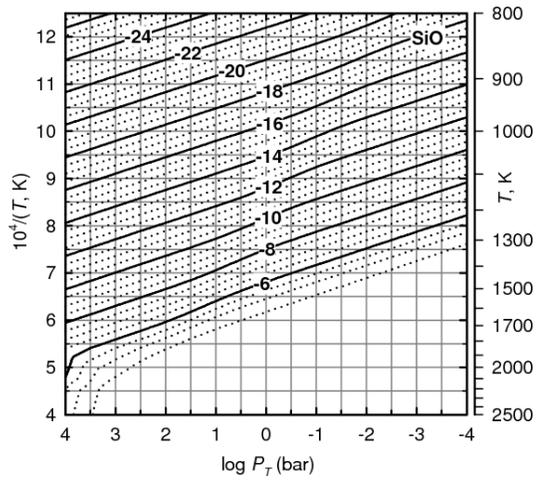}} \caption[Individual Si-bearing gases 1]{Mole fraction contours (on a logarithmic scale) for silicon monoxide (SiO) as a function of pressure and
temperature in a solar-metallicity gas.} \label{figure SiO}
\end{figure}

\begin{figure}
\scalebox{.7}{\includegraphics{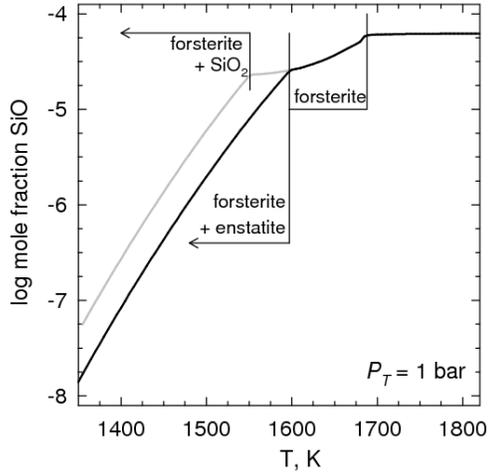}} \caption[Individual Si-bearing gases 1]{Mole fraction abundance of SiO gas at 1 bar total pressure in a solar-metallicity gas with enstatite formation
included (black line) or suppressed (gray line) above the forsterite cloud layer.  The formation of SiO$_{2}$ instead of MgSiO$_{3}$ would yield larger mole fraction abundances of SiO gas above
the clouds.} \label{figure SiO condense}
\end{figure}

\begin{figure}
\scalebox{.7}{\includegraphics{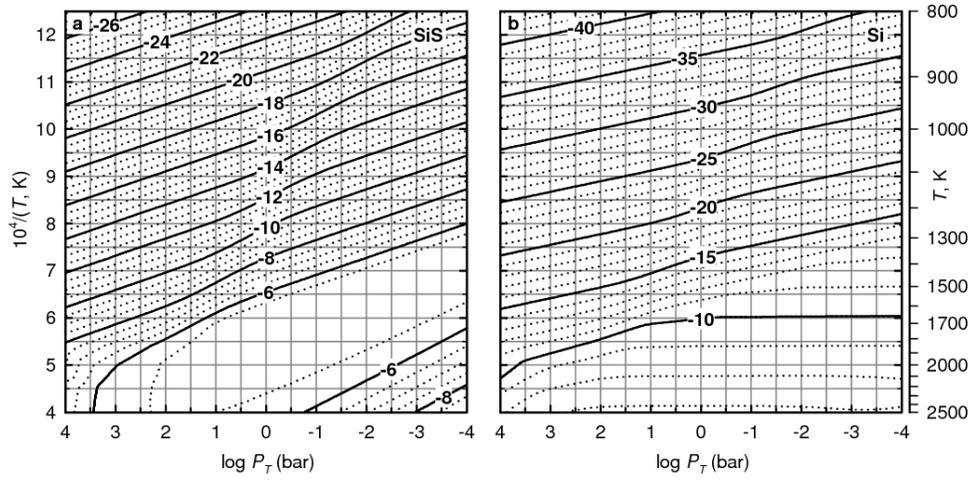}} \caption[Individual Si-bearing gases 1]{Mole fraction contours (on a logarithmic scale) for (a) silicon monosulfide (SiS) and and (b) monatomic silicon
(Si) as a function of pressure and temperature in a solar-metallicity gas.} \label{figure silicon gas 1}
\end{figure}

\begin{figure}
\scalebox{.7}{\includegraphics{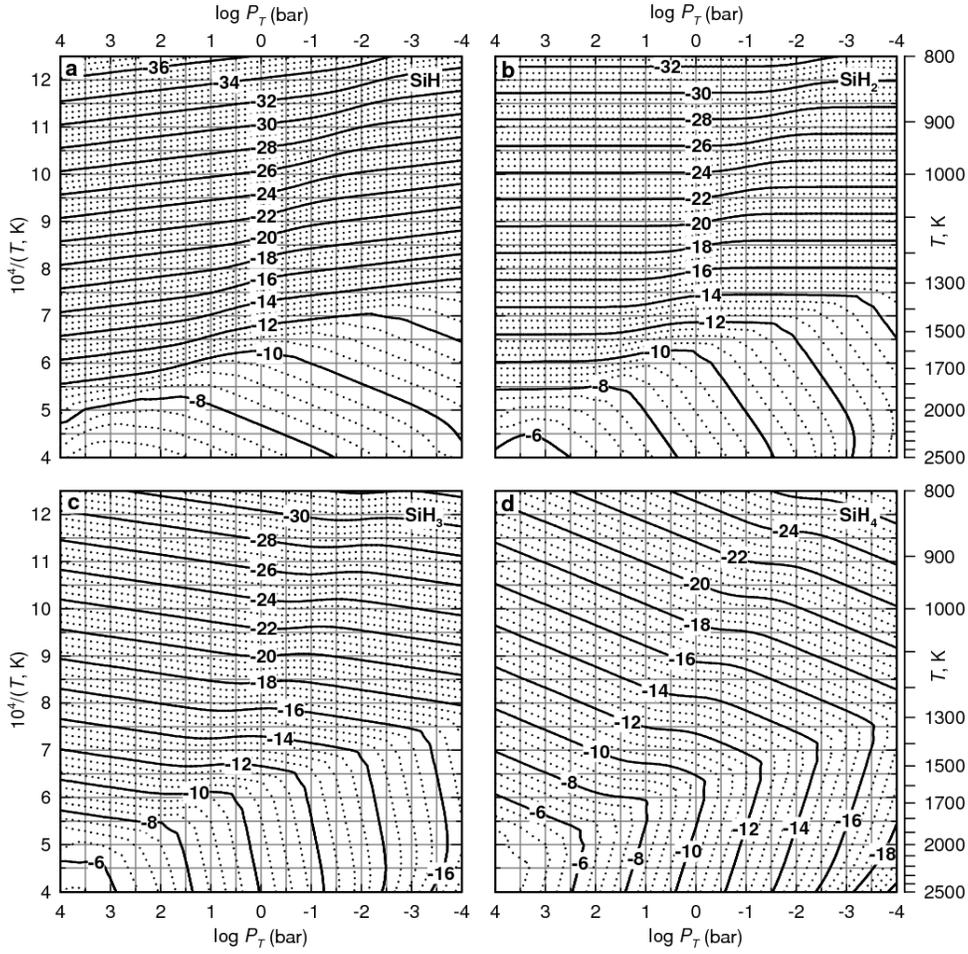}} \caption[Individual Si-bearing gases 2]{Mole fraction contours (on a logarithmic scale) for (a) silylidyne (SiH), (b) silylene (SiH$_{2}$), (c) the silyl
radical (SiH$_{3}$), and (d) silane (SiH$_{4}$) as a function of pressure and temperature in a solar-metallicity gas.  The bends in the abundance contours occur where magnesium silicate clouds
(MgSiO$_{3}$, Mg$_{2}$SiO$_{4}$) condense.} \label{figure silicon gas 2}
\end{figure}

\begin{figure}
\scalebox{.7}{\includegraphics{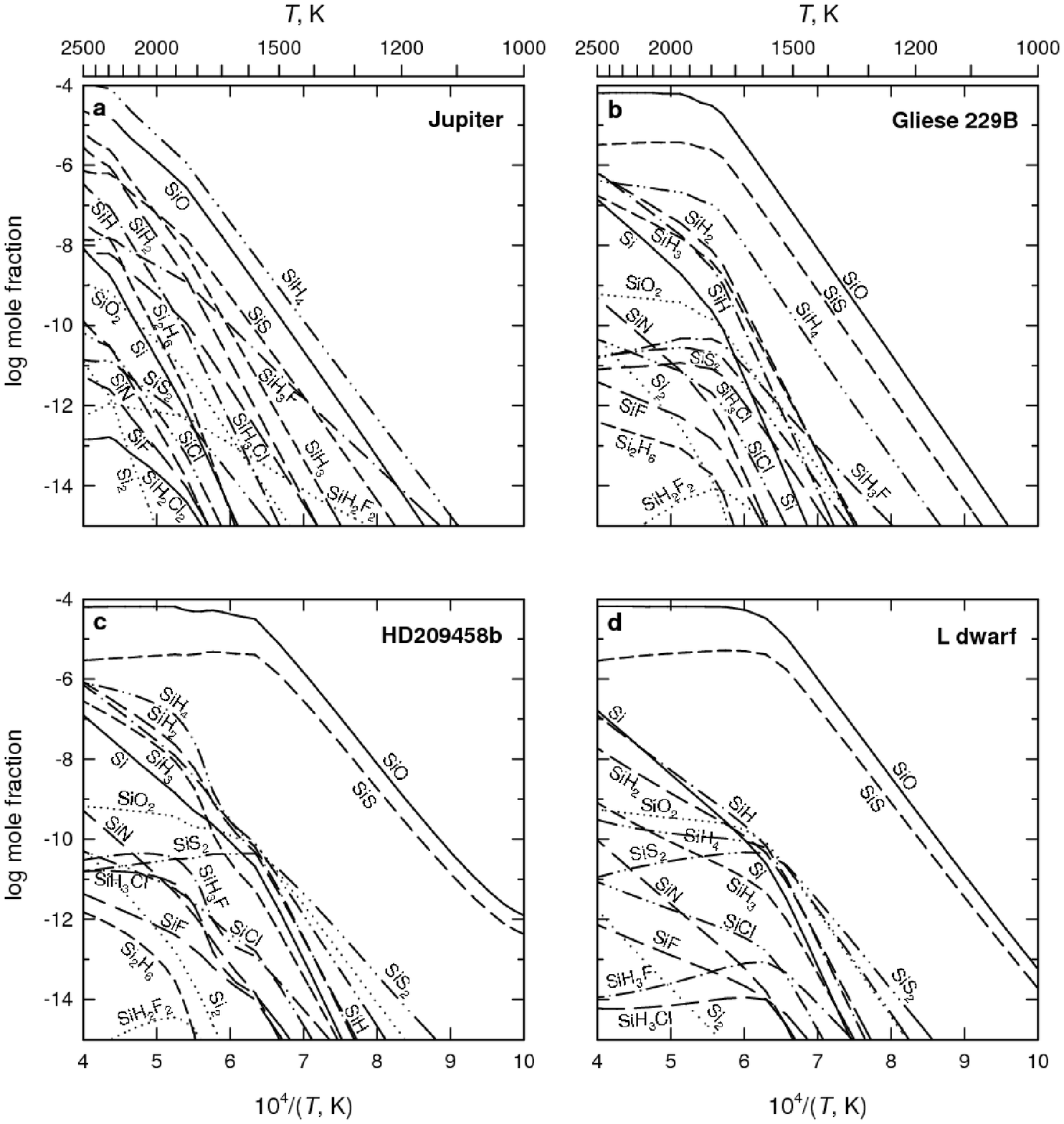}} \caption[Summary of silicon chemistry]{Silicon chemistry along the atmospheric pressure-temperature profile from 1000 to 2500 K for (a) Jupiter, (b)
Gliese 229B, (c) HD209458b, and (d) an L dwarf ($T_{\textrm{\scriptsize{eff}}}=1800$ K).  Breaks in the SiO curves show where magnesium-silicates condense (b,c,d).  Forsterite is condensed
throughout the entire temperature shown here for Jupiter (a).} \label{figure silicon chemistry summary}
\end{figure}

\end{document}